\documentclass[12pt,a4paper,english,aps,superscriptaddress,groupedaddress,showpacs,showkeys]{revtex4}

\usepackage[T1]{fontenc}
\usepackage[latin9]{inputenc}
\usepackage{babel}
\usepackage{amsmath}
\usepackage{amssymb}
\usepackage{esint}
\usepackage[unicode=true]
 {hyperref}
\usepackage{breakurl}

\makeatletter

\@ifundefined{textcolor}{}
{%
 \definecolor{BLACK}{gray}{0}
 \definecolor{WHITE}{gray}{1}
 \definecolor{RED}{rgb}{1,0,0}
 \definecolor{GREEN}{rgb}{0,1,0}
 \definecolor{BLUE}{rgb}{0,0,1}
 \definecolor{CYAN}{cmyk}{1,0,0,0}
 \definecolor{MAGENTA}{cmyk}{0,1,0,0}
 \definecolor{YELLOW}{cmyk}{0,0,1,0}
}
\newenvironment{lyxlist}[1]
{\begin{list}{}
{\settowidth{\labelwidth}{#1}
 \setlength{\leftmargin}{\labelwidth}
 \addtolength{\leftmargin}{\labelsep}
 }}
{\end{list}}

\usepackage{dcolumn}
\usepackage{bm}
\usepackage{amsfonts}

\makeatother

\begin{document}

\title{Nontrivial local observables and impermeable and permeable boundary
conditions for 1D KFGM particles}

\author{Techapon Kampu}

\homepage{https://orcid.org/0009-0003-1427-8409}

\email{[techaponk65@nu.ac.th]}

\affiliation{The Institute for Fundamental Study (IF), Naresuan University, Phitsanulok
65000, Thailand}

\author{Salvatore De Vincenzo}

\homepage{https://orcid.org/0000-0002-5009-053X}

\email{[salvatored@nu.ac.th]}

\affiliation{The Institute for Fundamental Study (IF), Naresuan University, Phitsanulok
65000, Thailand}

\thanks{S. De Vincenzo would like to dedicate this paper to the memory of
Luigi Mondino, physicist, teacher, tutor, and friend}

\date{December 23, 2025}

\begin{abstract}

\noindent Real solutions of the 1D Klein-Fock-Gordon (KFG) equation
automatically cancel out the usual two-vector current density; consequently,
the respective continuity equation is trivially satisfied, and a globally
conserved quantity cannot be obtained. Additionally, distinguishing
between impermeable and permeable boundary conditions (BCs) at a given
point is not possible. We address these first-quantized conflicts
by using the simplest nontrivial local observables, i.e., an energy
density and an energy current density, which allows us to characterize
a strictly neutral 1D KFG particle, i.e., a 1D KFG-Majorana (KFGM)
particle, when it is confined to an interval and when it is restricted,
e.g., in an interval with transparent walls. All the BCs for this
system are extracted from the pseudo self-adjointness of the Feshbach-Villars
(FV) Hamiltonian plus two Majorana conditions. We show that these
energy densities are components of an unusual energy-momentum tensor
and can satisfy a continuity equation, leading to a conserved quantity
for all available BCs. Moreover, the energy current density can characterize
all the BCs as either impermeable or permeable. In contrast, the commonly
used energy current density---a component of the usual energy-momentum
tensor---cannot characterize all the BCs. Additionally, this quantity
and its respective energy density---another component of the usual
energy-momentum tensor---may not lead to a conserved quantity. We
also obtain the BCs for which the abovementioned densities and those
commonly used are equally satisfactory. In fact, this occurs only
for four impermeable BCs and a one-parameter set of permeable BCs.
Our results highlight the important role played by the BCs when they
are imposed on a system in which particles occupy a finite region.
\end{abstract}

\pacs{03.65.-w, 03.65.Ca, 03.65.Db, 03.65.Pm}

\keywords{1D Klein-Fock-Gordon-Majorana particles; 1D Klein-Fock-Gordon wave
equation; 1D Feshbach-Villars wave equation; pseudo-Hermitian operator;
pseudo self-adjoint operator; local observables; quantum boundary
conditions; real scalar field}

\maketitle

\section{Introduction}

\noindent In (3+1) dimensions, there exist particles that are strictly
or intrinsically neutral, that is, particles that are their own antiparticles.
For these particles, all quantum numbers associated with electric
charge, such as the weak hypercharge, must be equal to zero. For example,
the neutral pion (spin-$0$) and the photon (spin-$1$) are strictly
neutral particles (although the neutral pion is a composite particle).
The graviton (spin-$2$) is also thought to be an example of such
a particle. The neutrino (spin-$\tfrac{1}{2}$) was initially considered
another example; however, each neutrino and its respective antiparticles
have different weak hypercharges. In any way, hypothetical spin-$\tfrac{1}{2}$
particles such as sterile neutrinos and neutralinos can be strictly
neutral particles; however, to date, no experimental evidence for
the existence of these particles has been found. For some years, spin-$\tfrac{1}{2}$
particles that are strictly neutral have been called Majorana particles,
i.e., Majorana fermions (because the Italian physicist Ettore Majorana,
first introduced this possibility by considering particles described
with the Dirac equation \cite{RefA}). However, we could also call
Majorana particles strictly neutral particles that have spins other
than $\tfrac{1}{2}$ because they also have a Majorana nature (see,
for example, Refs. \cite{RefB,RefC,RefD}). 

Recently, the specific case of one-dimensional (1D) Klein-Fock-Gordon
(KFG) particles, which are also strictly neutral and exist in one
space (a finite interval) and one time dimension, was studied. These
particles were referred to as 1D Klein-Fock-Gordon-Majorana (KFGM)
particles \cite{RefE}. Such first-quantized particles are described
by the standard 1D KFG and/or 1D Feshbach-Villars (FV) wave equations
\cite{RefF,RefG,RefH,RefI,RefJ}, each with a real Lorentz scalar
potential and some kind of Majorana condition. As shown in Ref. \cite{RefE},
certain first-order equations in the time derivative that do not have
a Hamiltonian form can also describe these particles. These equations
were called first-order 1D Majorana equations. 

For example, if we consider the 1D KFG equation, namely,
\begin{equation}
\hat{\mathrm{O}}_{\mathrm{KFG}}\psi\equiv\left[\,\hat{\mathrm{E}}^{2}-(c\,\hat{\mathrm{p}})^{2}-(\mathrm{m}c^{2})^{2}-2\,\mathrm{m}c^{2}S\,\right]\psi=0,
\end{equation}
where $\psi=\psi(x,t)$ is a one-component wavefunction, $\hat{\mathrm{E}}=\mathrm{i}\hbar\,\partial/\partial t$
is the energy operator, $\hat{\mathrm{p}}=-\mathrm{i}\hbar\,\partial/\partial x$
is the momentum operator, and $S=S(x,t)$ is a real-valued Lorentz
scalar interaction, the Majorana condition can be the reality condition
for the wavefunction $\psi$, i.e., $\psi=\psi^{*}$, but it can also
be $\psi=-\psi^{*}$ (the superscript $^{*}$ denotes the complex
conjugate, as usual), i.e., the wavefunction $\psi$ can be purely
imaginary \cite{RefE}. Unfortunately, as can be immediately verified,
the usual local observables or densities
\begin{equation}
\varrho=\varrho(x,t)=\frac{1}{2\mathrm{m}c^{2}}[\,\psi^{*}(\hat{\mathrm{E}}\psi)-(\hat{\mathrm{E}}\psi^{*})\,\psi\,]=\frac{1}{\mathrm{m}c^{2}}\,\mathrm{Re}[\,\psi^{*}(\hat{\mathrm{E}}\psi)\,]
\end{equation}
and
\begin{equation}
j=j(x,t)=\frac{1}{2\mathrm{m}c}\left[\,\psi^{*}(c\hat{\mathrm{p}}\psi)-(c\hat{\mathrm{p}}\psi^{*})\,\psi\,\right]=\frac{1}{\mathrm{m}c}\,\mathrm{Re}[\,\psi^{*}(c\hat{\mathrm{p}}\psi)\,],
\end{equation}
which must satisfy a continuity equation, i.e., $\hat{\mathrm{E}}\varrho-\hat{\mathrm{p}}j=0$,
leading to the result $\int_{\Omega}\mathrm{d}x\,\varrho=\mathrm{const}$
($\Omega=[a,b]$ is the interval where the particle is present), are
automatically zero when $\psi=\psi^{*}$ and $\psi=-\psi^{*}$, i.e.,
\begin{equation}
\varrho=\varrho(x,t)=0\quad\mathrm{and}\quad j=j(x,t)=0,
\end{equation}
and the continuity equation is trivially satisfied, and the result
$\int_{\Omega}\mathrm{d}x\,\varrho=\mathrm{const}$ cannot be obtained.
The latter is a well-known result in relativistic quantum mechanics. 

Additionally, when the solutions $\psi$ of the KFG equation are definitely
real, they always lead to so-called impenetrability condition at the
extremes of a finite interval $\Omega=[a,b]$. i.e., $j(b,t)=j(a,t)=0$.
Certainly, the solutions of the 1D KFG equation in Eq. (1) can be
chosen to be real; however, they need not be real, i.e., complex solutions
can also be written (we do not refer to imaginary solutions). This
situation arises because the operator $\hat{\mathrm{O}}_{\mathrm{KFG}}$
in Eq. (1) satisfies $(\hat{\mathrm{O}}_{\mathrm{KFG}}\psi)^{*}=\hat{\mathrm{O}}_{\mathrm{KFG}}\psi^{*}$,
i.e., $\hat{\mathrm{O}}_{\mathrm{KFG}}$ is a real operator. Thus,
when $\psi\in\mathbb{C}$, we can recognize impermeable (or confining
or impenetrable) boundary conditions (BCs) and permeable (or nonconfining
or penetrable or transparent) BCs \cite{RefK}. As a general rule,
confining BCs satisfy $j(b,t)=j(a,t)=0$, and nonconfining BCs satisfy
$j(b,t)=j(a,t)$ (for which, in the latter case, the solutions $\psi$
must be complex).

Thus, when the solutions of Eq. (1) are real-valued functions, that
is, when the solutions of the equation describe a 1D KFGM particle,
the distinction between confining and nonconfining BCs is not realizable,
at least if the current density $j$ is considered. Clearly, the introduction
of these two types of BCs would require the use of other current densities.
The latter issue was explicitly identified in Ref. \cite{RefE} but
was not solved. Thus, one of the goals of our work is to propose a
nontrivial local observable---a current density---that allows us to
completely characterize a 1D KFGM particle when it is confined to
an interval and when it is, say, in a box with transparent walls (or
walls that allow leaks). We also wish to report the specific BCs that
determine these two physical situations and examine the possibility
of obtaining a continuity equation involving this current density
and an additional density. Incidentally, the need to use local observables
other than $j$ and $\varrho$ when the 1D KFG equation describes
a strictly neutral particle has already been briefly mentioned in
the renowned paper by Feshbach and Villars \cite{RefI}. However,
to our knowledge, our work is the first to address the problem of
introducing confining and nonconfining BCs using local observables
that do not trivially vanish and are real quantities when the wavefunction
$\psi$ describes a 1D KFGM particle.

Our paper follows the following plan. In Section II, we introduce
the pseudo self-adjoint Hamiltonian operator that is present in the
1D FV wave equation. We also discuss certain general concepts within
operator theory that apply to this Hamiltonian and the consequences
of its pseudo self-adjointness. We also introduce here the most general
set of quantum BCs for which this Hamiltonian is pseudo self-adjoint
on a finite interval. If the solutions of the 1D KFG equations are
complex, this set depends on four real parameters. 

In Section III, we introduce three energy densities and two energy
current densities and relate them to the usual density $\varrho$
and current density $j$. We find that, given a BC, only one of the
energy current densities always has the same value at the extremes
of the interval, which is a consequence of the pseudo self-adjointness
of the FV Hamiltonian. In addition, this energy current density and
its associated energy density satisfy a continuity equation when the
Lorentz scalar potential does not explicitly depend on time (this
situation is valid on-shell). However, these two local observables
are complex quantities when the wavefunction $\psi$ is complex but
are real quantities if $\psi$ satisfies a Majorana condition, and
better yet, they are not automatically zero. We also find that the
spatial integral of the energy density, i.e., the expectation value
of the energy operator, can even be a positive quantity. The latter
statement essentially depends on the BCs and the positivity of a (time-independent)
scalar potential in the interval. On the other hand, the other energy
current density---the component $c\, T_{\;\;0}^{1}$ of the standard
momentum-energy tensor for this two-dimensional space time system---and
either of the other two energy densities---one of which is the $T_{\;\;0}^{0}$
component of $T_{\;\;\nu}^{\mu}$---are always real quantities; however,
given a BC, $c\, T_{\;\;0}^{1}$ does not necessarily have the same
value at the ends of the interval. The latter statement also depends
on the BCs.

Now, let us explain what we do in Section IV. First, the results presented
in Section III indicate that the abovementioned pair of local observables,
which are real quantities when the wavefunction $\psi$ obeys a Majorana
condition, seems to be acceptable for completely characterizing a
1D KFGM particle in an interval. On the other hand, the usual pair
of local observables $\{T_{\;\;0}^{0},c\, T_{\;\;0}^{1}\}$ seems
to be limited. In fact, throughout Section IV, we present results
specifically related to BCs that confirm this observation. In addition,
we present the three-parameter general family of BCs that results
from imposing a Majorana condition for the four-parameter general
set of BCs. We also obtain a condition that defines the confining
BCs for the 1D KFGM particles. Then, we write the most general two-parameter
subset of confining BCs for this class of particles. 

In this section, we also obtain the mathematical conditions that allow
the energy current density $c\, T_{\;\;0}^{1}$ to have the same value
at $x=a$ and $x=b$. We find that these are also the conditions that
would allow the mean value of the energy operator to definitely be
a positive quantity; however, in addition, this mean value would be
equal to the spatial integrals of the ``acceptable'' energy density
and $T_{\;\;0}^{0}$; thus, these two quantities would be equally
satisfactory, even though they always differ in the gradient of a
function. We show that only four typical impenetrable BCs verify these
mathematical conditions, namely, the Dirichlet and Neumann BCs and
two mixed BCs. In addition, a one-parameter set of nonconfining BCs
also verifies them. In all these cases, the two energy current densities
also appear to be equally satisfactory. For example, they satisfy
the same condition at the interval walls, although they also differ
in the gradient of a function. For all other BCs that do not satisfy
the mathematical conditions referred to here, only the ``acceptable''
energy current density can characterize the impenetrability and eventual
penetrability of the interval.

In Section IV, we also obtain the mathematical conditions that ensure
that the spatial integrals of the ``acceptable'' energy current
density and $c\, T_{\;\;0}^{1}$ are equal. In this case, we show
that the Dirichlet BC is the only confining BC that verifies these
conditions. Moreover, the BCs within a one-parameter set of nonconfining
BCs also satisfy these requirements. Finally, we discuss our results
in Section V, and results that complement what has been stated throughout
the article are presented in four appendices. In particular, in Appendix
D, we show that the two ``acceptable'' local observables are two
components of a rank-2 Lorentz tensor, an unusual energy-momentum
tensor, and satisfy a continuity equation when the scalar potential
is explicitly independent of time.

\section{Preliminaries}

\noindent Let us consider the Hamiltonian operator for the 1D FV wave
equation, i.e.,
\begin{equation}
\hat{\mathrm{h}}=\frac{\hat{\mathrm{p}}^{2}}{2\mathrm{m}}(\hat{\tau}_{3}+\mathrm{i}\hat{\tau}_{2})+\mathrm{m}c^{2}\hat{\tau}_{3}+S\,(\hat{\tau}_{3}+\mathrm{i}\hat{\tau}_{2}),
\end{equation}
where $\hat{\tau}_{3}=\hat{\sigma}_{z}$ and $\hat{\tau}_{2}=\hat{\sigma}_{y}$
are Pauli matrices. The FV equation with this Hamiltonian, i.e., 
\begin{equation}
\hat{\mathrm{O}}_{\mathrm{FV}}\Psi\equiv(\hat{\mathrm{E}}\,\hat{1}_{2}-\hat{\mathrm{h}})\Psi=0
\end{equation}
($\hat{1}_{2}$ is the $2\times2$ identity matrix), can describe
a strictly neutral 1D KFG particle, i.e., a 1D KFGM particle, but
in addition, a Majorana condition must hold \cite{RefE}. The operator
$\hat{\mathrm{h}}$ acts on two-component wavefunctions of the form
$\Psi=\Psi(x,t)=\left[\,\psi_{1}\;\,\psi_{2}\,\right]^{\mathrm{T}}=\left[\,\psi_{1}(x,t)\;\,\psi_{2}(x,t)\,\right]^{\mathrm{T}}$
(the symbol $^{\mathrm{T}}$ represents the transpose of a matrix).
The generalized Hermitian conjugate or the formal generalized adjoint
of $\hat{\mathrm{h}}$, that is, $\hat{\mathrm{h}}_{\mathrm{adj}}$,
is given by
\begin{equation}
\hat{\mathrm{h}}_{\mathrm{adj}}\equiv\hat{\tau}_{3}\,\hat{\mathrm{h}}^{\dagger}\,\hat{\tau}_{3}=\frac{\hat{\mathrm{p}}^{2}}{2\mathrm{m}}(\hat{\tau}_{3}+\mathrm{i}\hat{\tau}_{2})+\mathrm{m}c^{2}\hat{\tau}_{3}+S\,(\hat{\tau}_{3}+\mathrm{i}\hat{\tau}_{2})
\end{equation}
(the symbol $^{\mathrm{\dagger}}$ denotes the usual Hermitian conjugate
of a matrix and an operator; hence, $\hat{\mathrm{h}}_{\mathrm{adj}}$
and $\hat{\mathrm{h}}^{\dagger}$ are unitarily equivalent). Thus,
the action of $\hat{\mathrm{h}}_{\mathrm{adj}}$ is clearly the same
as that of $\hat{\mathrm{h}}$. The definition given in Eq. (7) is
equivalent to the following relation: 
\begin{equation}
\langle\langle\hat{\mathrm{h}}_{\mathrm{adj}}\Psi,\Phi\rangle\rangle=\langle\langle\Psi,\hat{\mathrm{h}}\Phi\rangle\rangle
\end{equation}
(recall that $\hat{\tau}_{3}=\hat{\tau}_{3}^{\dagger}=\hat{\tau}_{3}^{-1}$),
where the inner product is defined by
\begin{equation}
\langle\langle\Psi,\Phi\rangle\rangle\equiv\int_{\Omega}\mathrm{d}x\,\Psi^{\dagger}\hat{\tau}_{3}\Phi=\langle\langle\Phi,\Psi\rangle\rangle^{*}
\end{equation}
($\Psi=\left[\,\psi_{1}\;\,\psi_{2}\,\right]^{\mathrm{T}}$ and $\Phi=\left[\,\phi_{1}\;\,\phi_{2}\,\right]^{\mathrm{T}}$,
and the superscript $^{*}$ denotes the complex conjugate). The relation
in Eq. (8) defines the generalized Hermitian conjugate or the generalized
adjoint $\hat{\mathrm{h}}_{\mathrm{adj}}$ on an indefinite inner
product space; in our case, a $\hat{\tau}_{3}$-space \cite{RefL}.
Moreover, $\hat{\mathrm{h}}$ can act only on a specific set of functions
$\Phi$, the domain of $\hat{\mathrm{h}}$, i.e., $\mathrm{\mathcal{D}}(\hat{\mathrm{h}})$;
and $\hat{\mathrm{h}}_{\mathrm{adj}}$ can act only on the set of
functions $\Psi$ in its domain $\mathrm{\mathcal{D}}(\hat{\mathrm{h}}_{\mathrm{adj}})$.
We also assume that $\mathrm{\mathcal{D}}(\hat{\mathrm{h}})$ is weakly
dense in space; hence, $\hat{\mathrm{h}}_{\mathrm{adj}}$ is unique.
The scalar product in Eq. (9), which is---an indefinite (or improper)
inner product---, becomes a real number---positive and even negative---when
$\Phi=\Psi$. In the first-quantized KFG theory, the latter is not
a serious problem; in fact, the scalar product $\langle\langle\Psi,\Psi\rangle\rangle$
is the integral of the usual density $\varrho$, and it can ultimately
be interpreted as a charge density. 

In general, an indefinite inner product space---an infinite-dimensional
complex vector space---always comes with an indefinite scalar product,
such as the one given in Eq. (9), and with a positive semidefinite
inner product. In our case, the latter inner product is the usual
scalar product of Dirac theory in one spatial dimension, namely, $\langle\Psi,\Phi\rangle_{\mathrm{D}}\equiv\langle\langle\Psi,\hat{\tau}_{3}\Phi\rangle\rangle$
(recall that $\hat{\tau}_{3}^{2}=\hat{1}_{2}$, i.e., $\hat{\tau}_{3}^{-1}=\hat{\tau}_{3}$);
hence, $\langle\langle\Psi,\Phi\rangle\rangle=\langle\Psi,\hat{\tau}_{3}\Phi\rangle_{\mathrm{D}}$
(see Eq. (9)). As we know, $\langle\Psi,\Phi\rangle_{\mathrm{D}}$
is a positive definite scalar product, and we say that the indefinite
inner product space is a Krein space (subject to the existence of
a locally convex topology where the inner product is jointly continuous)
\cite{RefL,RefM}.

We can say that $\hat{\mathrm{h}}$ in Eq. (5) is formally pseudo-Hermitian
or formally generalized Hermitian. Here, the word ``formally'' refers
to the fact that the actions of $\hat{\mathrm{h}}$ and $\hat{\mathrm{h}}_{\mathrm{adj}}=\hat{\tau}_{3}\,\hat{\mathrm{h}}^{\dagger}\,\hat{\tau}_{3}$
are equal, and additionally, we are unconcerned with the type of functions
on which these two operators can act, i.e., with the domains of the
operators. However, BCs are an essential part of any unbounded operator. 

Indeed, by applying the method of integration by parts twice, we can
check that $\hat{\mathrm{h}}$ and $\hat{\mathrm{h}}_{\mathrm{adj}}$
(which acts as $\hat{\mathrm{h}}$) satisfy the following relation
\cite{RefK}: 
\begin{equation}
\langle\langle\hat{\mathrm{h}}_{\mathrm{adj}}\Psi,\Phi\rangle\rangle=\langle\langle\Psi,\hat{\mathrm{h}}\Phi\rangle\rangle-\frac{\hbar^{2}}{2\mathrm{m}}\,\frac{1}{2}\left.\left[\,\left((\hat{\tau}_{3}+\mathrm{i}\hat{\tau}_{2})\Psi_{x}\right)^{\dagger}(\hat{\tau}_{3}+\mathrm{i}\hat{\tau}_{2})\Phi-\left((\hat{\tau}_{3}+\mathrm{i}\hat{\tau}_{2})\Psi\right)^{\dagger}(\hat{\tau}_{3}+\mathrm{i}\hat{\tau}_{2})\Phi_{x}\,\right]\right|_{a}^{b},
\end{equation}
where $\Psi\in\mathrm{\mathcal{D}}(\hat{\mathrm{h}}_{\mathrm{adj}})$
and $\Phi\in\mathrm{\mathcal{D}}(\hat{\mathrm{h}})$, $\left.\left[\, f\,\right]\right|_{a}^{b}\equiv f(b,t)-f(a,t)$,
and $\Psi_{x}\equiv\partial\Psi/\partial x$, etc. If the BCs imposed
on $\Psi$ and $\Phi$ at the ends of the interval $\Omega$ lead
to the disappearance of the boundary term in Eq. (10), the operator
$\hat{\mathrm{h}}$ would be (more than just formally) pseudo-Hermitian
or generalized Hermitian. These BCs do not have to be the same, i.e.,
the domains $\mathrm{\mathcal{D}}(\hat{\mathrm{h}}_{\mathrm{adj}})$
and $\mathrm{\mathcal{D}}(\hat{\mathrm{h}})$ do not have to be identical.
In fact, one might not have to impose a specific BC on $\Psi$, although
one would have to impose one on $\Phi$. For example, inside $\mathrm{\mathcal{D}}(\hat{\mathrm{h}})$,
we choose $(\hat{\tau}_{3}+\mathrm{i}\hat{\tau}_{2})\Phi(a,t)=(\hat{\tau}_{3}+\mathrm{i}\hat{\tau}_{2})\Phi(b,t)=0$,
and $(\hat{\tau}_{3}+\mathrm{i}\hat{\tau}_{2})\Phi_{x}(a,t)=(\hat{\tau}_{3}+\mathrm{i}\hat{\tau}_{2})\Phi_{x}(b,t)=0$.
Thus, in this case, the cancellation of the boundary term in Eq. (10)
does not require the presence of a BC inside $\mathrm{\mathcal{D}}(\hat{\mathrm{h}}_{\mathrm{adj}})$,
i.e., $\mathrm{\mathcal{D}}(\hat{\mathrm{h}})\subset\mathrm{\mathcal{D}}(\hat{\mathrm{h}}_{\mathrm{adj}})$. 

Specifically, the most general set of BCs leading to cancellation
of the boundary term was obtained in Ref. \cite{RefK}. For all BCs
included in that set, $\hat{\mathrm{h}}$ is a pseudo-Hermitian operator,
but it is also a pseudo self-adjoint operator, i.e., 
\begin{equation}
\langle\langle\hat{\mathrm{h}}\Psi,\Phi\rangle\rangle=\langle\langle\Psi,\hat{\mathrm{h}}\Phi\rangle\rangle.
\end{equation}
Consequently, the domain of $\hat{\mathrm{h}}$ contains the same
BCs as the domain of its generalized adjoint $\hat{\mathrm{h}}_{\mathrm{adj}}$,
i.e., $\mathrm{\mathcal{D}}(\hat{\mathrm{h}}_{\mathrm{adj}})=\mathrm{\mathcal{D}}(\hat{\mathrm{h}})$,
and we can write the equality $\hat{\mathrm{h}}=\hat{\mathrm{h}}_{\mathrm{adj}}$.
Thus, the general set of BCs is found in both domains. As a consequence
of only the relation given in Eq. (11) (even considering this result
as a formal result), several properties that are similar to those
commonly obeyed by formal Hermitian operators in the usual sense are
obtained. For example, the generalized mean value of the Hamiltonian
operator is real valued, i.e., $\langle\langle\hat{\mathrm{h}}\rangle\rangle_{\Psi}\equiv\langle\langle\Psi,\hat{\mathrm{h}}\Psi\rangle\rangle=\langle\langle\hat{\mathrm{h}}\Psi,\Psi\rangle\rangle=\langle\langle\Psi,\hat{\mathrm{h}}\Psi\rangle\rangle^{*}$
\cite{RefJ,RefN,RefO}. 

The following relation is also verified: 
\begin{equation}
\frac{\mathrm{d}}{\mathrm{d}t}\langle\langle\Psi,\Phi\rangle\rangle=\frac{1}{\mathrm{i}\hbar}\frac{\hbar^{2}}{2\mathrm{m}}\,\frac{1}{2}\left.\left[\,\left((\hat{\tau}_{3}+\mathrm{i}\hat{\tau}_{2})\Psi_{x}\right)^{\dagger}(\hat{\tau}_{3}+\mathrm{i}\hat{\tau}_{2})\Phi-\left((\hat{\tau}_{3}+\mathrm{i}\hat{\tau}_{2})\Psi\right)^{\dagger}(\hat{\tau}_{3}+\mathrm{i}\hat{\tau}_{2})\Phi_{x}\,\right]\right|_{a}^{b};
\end{equation}
however, $\Psi$ and $\Phi$ in Eq. (12) must satisfy the 1D FV equation.
Thus, the pseudo self-adjointness of $\hat{\mathrm{h}}$ implies that
the inner product defined in Eq. (9), namely, $\langle\langle\Psi,\Phi\rangle\rangle$,
is a constant independent of time (compare Eq. (10), where $\hat{\mathrm{h}}=\hat{\mathrm{h}}_{\mathrm{adj}}$,
such that Eq. (11) is verified, with Eq. (12)). When $\hat{\mathrm{h}}=\hat{\mathrm{h}}_{\mathrm{adj}}$,
the boundary term in Eq. (10) vanishes; therefore, the boundary term
in Eq. (12) also vanishes (in fact, the boundary term in Eq. (10)
is $-\mathrm{i}\hbar$ times the boundary term in Eq. (12)). The relations
between the two-component wavefunctions $\Psi$ and $\Phi$ (which
are solutions of the 1D FV equation) and the respective one-component
wavefunctions $\psi$ and $\phi$ (which are solutions of the 1D KFG
equation) are given by 
\begin{equation}
\Psi=\left[\begin{array}{c}
\psi_{1}\\
\psi_{2}
\end{array}\right]=\frac{1}{2}\left[\begin{array}{c}
\psi+\frac{1}{\mathrm{m}c^{2}}\hat{\mathrm{E}}\psi\\
\psi-\frac{1}{\mathrm{m}c^{2}}\hat{\mathrm{E}}\psi
\end{array}\right]\quad\mathrm{and}\quad\Phi=\left[\begin{array}{c}
\phi_{1}\\
\phi_{2}
\end{array}\right]=\frac{1}{2}\left[\begin{array}{c}
\phi+\frac{1}{\mathrm{m}c^{2}}\hat{\mathrm{E}}\phi\\
\phi-\frac{1}{\mathrm{m}c^{2}}\hat{\mathrm{E}}\phi
\end{array}\right],
\end{equation}
which implies that  $\psi_{1}+\psi_{2}=\psi$, $\phi_{1}+\phi_{2}=\phi$,
and $\psi_{1}-\psi_{2}=(\hat{\mathrm{E}}\psi/\mathrm{m}c^{2})$, $\phi_{1}-\phi_{2}=(\hat{\mathrm{E}}\phi/\mathrm{m}c^{2})$.
The scalar product introduced in Eq. (9) can also be written as follows:
\begin{equation}
\langle\langle\Psi,\Phi\rangle\rangle=\frac{1}{2\mathrm{m}c^{2}}\int_{\Omega}\mathrm{d}x\left[\,\psi^{*}(\hat{\mathrm{E}}\phi)-(\hat{\mathrm{E}}\psi^{*})\,\phi\,\right].
\end{equation}

\noindent The latter result can be obtained by substituting the relations
given in Eq. (13) into the definition given in Eq. (9). Using the
formula given in Eq. (14), we obtain again the result given in Eq.
(12), but this time, the boundary term depends on the solutions $\psi$
and $\phi$, namely, 
\begin{equation}
\frac{\mathrm{d}}{\mathrm{d}t}\langle\langle\Psi,\Phi\rangle\rangle=-\frac{1}{2\mathrm{m}c}\left.\left[\,\psi^{*}(c\hat{\mathrm{p}}\phi)-(c\hat{\mathrm{p}}\psi^{*})\,\phi\,\right]\right|_{a}^{b}.
\end{equation}
We recall that Eqs. (12) and (15) are valid on-shell. Note that if
we make $\Psi=\Phi$ and $\psi=\phi$ in Eq. (15), we obtain the result
$\mathrm{d}\langle\langle\Psi,\Psi\rangle\rangle/\mathrm{d}t=-\left.\left[\, j\,\right]\right|_{a}^{b}$
(see Eq. (3)); that is, the pseudo self-adjointness of $\hat{\mathrm{h}}$
also implies that the current density calculated for any solution
of the equations of motion satisfies the relation
\begin{equation}
j(b,t)=j(a,t)\quad\Rightarrow\;\;\left.\mathrm{Re}[\,\psi^{*}(c\hat{\mathrm{p}}\psi)\,]\right|_{a}^{b}=0
\end{equation}
 (see Eq. (3)). In fact, this last result is an immediate consequence
only of the pseudo self-adjointness of $\hat{\mathrm{h}}$ (see Appendix
A).

For completeness, we write here the most general four-parameter BC
set for the solutions $\psi$ of the 1D KFG equation, for which the
operator $\hat{\mathrm{h}}$ in the 1D FV equation is pseudo self-adjoint,
namely,
\begin{equation}
\left[\begin{array}{c}
\psi(b,t)-\mathrm{i}\lambda\psi_{x}(b,t)\\
\psi(a,t)+\mathrm{i}\lambda\psi_{x}(a,t)
\end{array}\right]=\hat{\mathrm{U}}_{(2\times2)}\left[\begin{array}{c}
\psi(b,t)+\mathrm{i}\lambda\psi_{x}(b,t)\\
\psi(a,t)-\mathrm{i}\lambda\psi_{x}(a,t)
\end{array}\right],
\end{equation}
where $\lambda\in\mathbb{R}$ is a parameter inserted for dimensional
reasons, and the $2\times2$ unitary matrix $\hat{\mathrm{U}}_{(2\times2)}$
is given by
\begin{equation}
\hat{\mathrm{U}}_{(2\times2)}=\mathrm{e}^{\mathrm{i}\,\mu}\left[\begin{array}{cc}
\mathrm{m}_{0}-\mathrm{i}\,\mathrm{m}_{3} & -\mathrm{m}_{2}-\mathrm{i}\,\mathrm{m}_{1}\\
\mathrm{m}_{2}-\mathrm{i}\,\mathrm{m}_{1} & \mathrm{m}_{0}+\mathrm{i}\,\mathrm{m}_{3}
\end{array}\right],
\end{equation}
where $\mu\in[0,\pi)$, and the real quantities $\mathrm{m}_{0}$,
$\mathrm{m}_{1}$, $\mathrm{m}_{2}$ and $\mathrm{m}_{3}$ satisfy
$(\mathrm{m}_{0})^{2}+(\mathrm{m}_{1})^{2}+(\mathrm{m}_{2})^{2}+(\mathrm{m}_{3})^{2}=1$
(see Ref. \cite{RefK}). Similarly, we can report the following relation:
which gives us the explicit value of the current density given in
Eq. (3) at $x=a$, namely,
\begin{equation}
j(x=a,t)=-\frac{\hbar}{\mathrm{m}}\frac{1}{\lambda}\,\mathrm{Im}\left[\left(\frac{\mathrm{m}_{1}+\mathrm{i}\,\mathrm{m}_{2}}{\mathrm{m}_{0}+\cos(\mu)}\right)\psi^{*}(a,t)\,\psi(b,t)\,\right],
\end{equation}
which, by virtue of the pseudo self-adjointness of $\hat{\mathrm{h}}$,
is also equal to $j(x=b,t)$ (see Appendix B). Clearly, when we make
$\mathrm{m}_{1}=\mathrm{m}_{2}=0$ in Eq. (19), then $j(a,t)=j(b,t)=0$.
The latter condition defines the confining BCs specifically when the
wavefunction $\psi$ is complex. However, if the solutions $\psi$
are real or purely imaginary, this condition produces the result $\mathrm{m}_{2}=0$
in the general set of BCs given in Eq. (17) (because $\psi$ and $\psi^{*}$
satisfy Eq. (17)). Because the imaginary part of a real number is
zero, we obtain again the result $j(a,t)=j(b,t)=0$ from Eq. (19),
which is expected. In Appendix B, we explicitly prove that all the
confining BCs for the (complex) solutions of the 1D KFG equation lead
immediately to the relation $j(a,t)=j(b,t)=0$.

\section{Nontrivial local observables}

\noindent Similar to the relation given in Eq. (10), the following
relation is also satisfied by $\hat{\mathrm{h}}$ and $\hat{\mathrm{h}}_{\mathrm{adj}}$:
\begin{equation}
\langle\langle\hat{\mathrm{h}}_{\mathrm{adj}}\Psi,\hat{\mathrm{E}}\Phi\rangle\rangle=\langle\langle\Psi,\hat{\mathrm{h}}\,\hat{\mathrm{E}}\Phi\rangle\rangle-\mathrm{i}\hbar\,\frac{\hbar^{2}}{2\mathrm{m}}\,\frac{1}{2}\left.\left[\,\left((\hat{\tau}_{3}+\mathrm{i}\hat{\tau}_{2})\Psi_{x}\right)^{\dagger}(\hat{\tau}_{3}+\mathrm{i}\hat{\tau}_{2})\dot{\Phi}-\left((\hat{\tau}_{3}+\mathrm{i}\hat{\tau}_{2})\Psi\right)^{\dagger}(\hat{\tau}_{3}+\mathrm{i}\hat{\tau}_{2})\dot{\Phi}_{x}\,\right]\right|_{a}^{b},
\end{equation}
where $\dot{\Phi}=\partial\Phi/\partial t$, etc. If $\Phi\in\mathrm{\mathcal{D}}(\hat{\mathrm{h}})$,
then, automatically, $\hat{\mathrm{E}}\Phi=\mathrm{i}\hbar\dot{\Phi}\in\mathrm{\mathcal{D}}(\hat{\mathrm{h}})$.
If $\Phi$ satisfies a BC, then $\dot{\Phi}$ satisfies that same
BC. Consequently, any BC included in the most general set of BCs,
which we know leads to the cancellation of the boundary term in Eq.
(10), must also cancel out the boundary term in Eq. (20). We recall
that for any of the BCs that are within the general family of BCs,
$\hat{\mathrm{h}}$ is a pseudo self-adjoint operator, i.e., $\hat{\mathrm{h}}=\hat{\mathrm{h}}_{\mathrm{adj}}$
(i.e., $\hat{\mathrm{h}}$ acts the same as $\hat{\mathrm{h}}_{\mathrm{adj}}$
and $\mathrm{\mathcal{D}}(\hat{\mathrm{h}})=\mathrm{\mathcal{D}}(\hat{\mathrm{h}}_{\mathrm{adj}})$)
\cite{RefK}. 

The following relation is also verified: 
\[
\frac{\mathrm{d}}{\mathrm{d}t}\langle\langle\Psi,\hat{\mathrm{E}}\Phi\rangle\rangle=+\frac{\hbar^{2}}{2\mathrm{m}}\,\frac{1}{2}\left.\left[\,\left((\hat{\tau}_{3}+\mathrm{i}\hat{\tau}_{2})\Psi_{x}\right)^{\dagger}(\hat{\tau}_{3}+\mathrm{i}\hat{\tau}_{2})\dot{\Phi}-\left((\hat{\tau}_{3}+\mathrm{i}\hat{\tau}_{2})\Psi\right)^{\dagger}(\hat{\tau}_{3}+\mathrm{i}\hat{\tau}_{2})\dot{\Phi}_{x}\,\right]\right|_{a}^{b}
\]
\begin{equation}
+\frac{1}{2}\int_{\Omega}\mathrm{d}x\,\dot{S}\left((\hat{\tau}_{3}+\mathrm{i}\hat{\tau}_{2})\Psi\right)^{\dagger}(\hat{\tau}_{3}+\mathrm{i}\hat{\tau}_{2})\Phi,
\end{equation}
where $\dot{S}=\partial S/\partial t$; additionally, $\Psi$ and
$\Phi$ in Eq. (21) must satisfy the 1D FV equation. If $S$ does
not explicitly depend on time, then the pseudo self-adjointness of
$\hat{\mathrm{h}}$ implies that $\langle\langle\Psi,\hat{\mathrm{E}}\Phi\rangle\rangle$
is a time-independent constant. This implication emerges because the
boundary term in Eq. (20) is zero when $\hat{\mathrm{h}}=\hat{\mathrm{h}}_{\mathrm{adj}}$;
therefore, the boundary term in Eq. (21) is also zero (in fact, the
boundary term in Eq. (20) is $-\mathrm{i}\hbar$ times the boundary
term in Eq. (21)). Using the relations given in Eq. (13), the scalar
product $\langle\langle\Psi,\hat{\mathrm{E}}\Phi\rangle\rangle$ can
also be written as follows: 
\begin{equation}
\langle\langle\Psi,\hat{\mathrm{E}}\Phi\rangle\rangle=\frac{1}{2\mathrm{m}c^{2}}\int_{\Omega}\mathrm{d}x\left[\,\psi^{*}\left(\hat{\mathrm{E}}\,(\hat{\mathrm{E}}\phi)\right)-(\hat{\mathrm{E}}\psi^{*})\,(\hat{\mathrm{E}}\phi)\,\right].
\end{equation}
Similarly, taking the time derivative of this last expression, we
obtain another expression for the result given in Eq. (21), namely,
\begin{equation}
\frac{\mathrm{d}}{\mathrm{d}t}\langle\langle\Psi,\hat{\mathrm{E}}\Phi\rangle\rangle=-\frac{1}{2\mathrm{m}c}\left.\left[\,\psi^{*}\left(c\hat{\mathrm{p}}\,(\hat{\mathrm{E}}\phi)\right)-(c\hat{\mathrm{p}}\psi^{*})\,(\hat{\mathrm{E}}\phi)\,\right]\right|_{a}^{b}+\int_{\Omega}\mathrm{d}x\,\dot{S}\,\psi^{*}\phi,
\end{equation}
where $\psi$ and $\phi$ satisfy the 1D KFG equation. If we make
$\Psi=\Phi$ and $\psi=\phi$ in Eq. (23), we obtain the result
\begin{equation}
\frac{\mathrm{d}}{\mathrm{d}t}\langle\langle\Psi,\hat{\mathrm{E}}\Psi\rangle\rangle=\frac{\mathrm{d}}{\mathrm{d}t}\left(\int_{\Omega}\mathrm{d}x\,\varrho_{\mathrm{E}}\right)=-\left.\left[\, j_{\mathrm{E}}\,\right]\right|_{a}^{b}+\int_{\Omega}\mathrm{d}x\,\dot{S}\,\psi^{*}\psi,
\end{equation}
where 
\begin{equation}
\varrho_{\mathrm{E}}=\varrho_{\mathrm{E}}(x,t)=\frac{1}{2\mathrm{m}c^{2}}\left[\,\psi^{*}\left(\hat{\mathrm{E}}\,(\hat{\mathrm{E}}\psi)\right)-(\hat{\mathrm{E}}\psi^{*})\,(\hat{\mathrm{E}}\psi)\,\right]
\end{equation}
is a type of complex energy density \cite{RefO}, and
\begin{equation}
j_{\mathrm{E}}=j_{\mathrm{E}}(x,t)=\frac{1}{2\mathrm{m}c}\left[\,\psi^{*}\left(c\hat{\mathrm{p}}\,(\hat{\mathrm{E}}\psi)\right)-(c\hat{\mathrm{p}}\psi^{*})\,(\hat{\mathrm{E}}\psi)\,\right]
\end{equation}
is a type of complex energy current density. These two quantities
are complex only when $\psi$ is a complex wavefunction; thus, in
this case, they may be physically unsuitable unless integrating them
in the interval $\Omega$ yields a real quantity. Similarly, if we
make $\Psi=\Phi$ in Eq. (20) (and then $\psi=\phi$), we obtain the
result
\begin{equation}
j_{\mathrm{E}}(b,t)=j_{\mathrm{E}}(a,t),
\end{equation}
which is a straightforward consequence of the pseudo self-adjointness
of $\hat{\mathrm{h}}$. Clearly, the result in Eq. (27) implies a
zero accumulation of energy flow at the ends of the interval. The
densities $\varrho_{\mathrm{E}}$ and $j_{\mathrm{E}}$ can be rewritten
in a form that allows us to relate them to the usual densities, new
densities and other quantities, namely,
\begin{equation}
\varrho_{\mathrm{E}}=-\frac{\mathrm{i}}{2\mathrm{m}c^{2}}\,\hat{\mathrm{E}}\,[\,\mathrm{Im}(\psi^{*}\,(\hat{\mathrm{E}}\psi))\,]+\frac{1}{2}\,\hat{\mathrm{E}}\,\varrho+\tilde{\varrho}_{\mathrm{E}},
\end{equation}
where 
\begin{equation}
\tilde{\varrho}_{\mathrm{E}}=\frac{1}{2\mathrm{m}c^{2}}\left[\,\psi^{*}(\hat{\mathrm{E}}^{2}\psi)+(\hat{\mathrm{E}}^{2}\psi^{*})\,\psi\,\right]=\frac{1}{\mathrm{m}c^{2}}\,\mathrm{Re}\left[\,\psi^{*}(\hat{\mathrm{E}}^{2}\psi)\,\right]
\end{equation}
is a real energy density. Alternatively, using the KFG equation (Eq.
(1)) and the continuity equation $\hat{\mathrm{E}}\varrho-\hat{\mathrm{p}}j=0$,
we can write the relation in Eq. (28) as follows:
\begin{equation}
\varrho_{\mathrm{E}}=+\frac{\mathrm{i}}{2\mathrm{m}c^{2}}\, c\hat{\mathrm{p}}\,[\,\mathrm{Im}(\psi^{*}\,(c\hat{\mathrm{p}}\psi))\,]+\frac{1}{2}\,\hat{\mathrm{E}}\,\varrho+T_{\;\;0}^{0},
\end{equation}
where 
\begin{equation}
T_{\;\;0}^{0}=\frac{1}{2\mathrm{m}c^{2}}\left[\,-(\hat{\mathrm{E}}\psi^{*})(\hat{\mathrm{E}}\psi)-(c\hat{\mathrm{p}}\psi^{*})(c\hat{\mathrm{p}}\psi)+(\mathrm{m}c^{2})^{2}\,\psi^{*}\psi+2\,\mathrm{m}c^{2}S\,\psi^{*}\psi\,\right]
\end{equation}
is another real energy density (and positive when $S\geq0$ in $\Omega$);
in fact, it is a component of the usual energy-momentum tensor of
the KFG theory in (1+1) dimensions (see Appendix C and, for example,
Ref. \cite{RefP}). Equations (28) and (30) clearly show that when
the wavefunction $\psi$ is complex, $\varrho_{\mathrm{E}}$ is complex
because $\hat{\mathrm{E}}\,\varrho$ is complex ($\hat{\mathrm{E}}\,\varrho$
is purely imaginary); however, if $\psi$ is real, $\hat{\mathrm{E}}\,\varrho=0$
and $\varrho_{\mathrm{E}}$ is real, which is physically adequate.
Likewise,
\begin{equation}
j_{\mathrm{E}}=\frac{\mathrm{i}}{2\mathrm{m}c}\,\hat{\mathrm{E}}\left[\,\mathrm{Im}(\psi^{*}\,(c\hat{\mathrm{p}}\psi))\,\right]+\frac{1}{2}\,\hat{\mathrm{E}}\, j+\tilde{j}_{\mathrm{E}},
\end{equation}
where
\begin{equation}
\tilde{j}_{\mathrm{E}}\equiv c\, T_{\;\;0}^{1}=-\frac{1}{2\mathrm{m}c}[\,(\hat{\mathrm{E}}\psi^{*})\,(c\hat{\mathrm{p}}\psi)+(c\hat{\mathrm{p}}\psi^{*})\,(\hat{\mathrm{E}}\psi)\,]=-\frac{1}{\mathrm{m}c}\,\mathrm{Re}[\,(\hat{\mathrm{E}}\psi^{*})\,(c\hat{\mathrm{p}}\psi)\,]
\end{equation}
is a real energy current density that is proportional to the component
$T_{\;\;0}^{1}$ of the usual energy-momentum tensor in the 1D KFG
theory (see Appendix C and Ref. \cite{RefP}). Additionally, when
the wavefunction $\psi$ is complex, $j_{\mathrm{E}}$ in Eq. (32)
is complex because $\hat{\mathrm{E}}\, j$ is complex (in fact, $\hat{\mathrm{E}}\, j$
is purely imaginary); however, if $\psi$ is real, $\hat{\mathrm{E}}\, j=0$
and $j_{\mathrm{E}}$ is real, which is physically appropriate. 

Then, as a consequence of the pseudo self-adjointness of $\hat{\mathrm{h}}$
(i.e., $[\,\hat{\mathrm{E}}j\,]{}_{a}^{b}=\hat{\mathrm{E}}\,[\, j\,]{}_{a}^{b}=0$
and $[\, j_{\mathrm{E}}\,]{}_{a}^{b}=0$), the result in Eq. (32)
implies that
\begin{equation}
0=\frac{\mathrm{i}}{2\mathrm{m}c}\left.\hat{\mathrm{E}}\left[\,\mathrm{Im}(\psi^{*}\,(c\hat{\mathrm{p}}\psi))\,\right]\right|_{a}^{b}+\left.\left[\,\tilde{j}_{\mathrm{E}}\,\right]\right|_{a}^{b}.
\end{equation}
Thus, if the first term evaluated at the boundaries in Eq. (34) vanishes
(which depends on the BCs), then it follows that
\begin{equation}
\tilde{j}_{\mathrm{E}}(b,t)=\tilde{j}_{\mathrm{E}}(a,t).
\end{equation}
Unfortunately, not all available BCs for this problem lead to the
disappearance of the boundary term $\tfrac{\mathrm{i}}{2\mathrm{m}c}\left.\hat{\mathrm{E}}\left[\,\mathrm{Im}(\psi^{*}\,(c\hat{\mathrm{p}}\psi))\,\right]\right|_{a}^{b}$
in Eq. (34), that is, to the verification of Eq. (35). In the next
section, we address this issue by specifically describing a 1D KFGM
particle. In Ref. \cite{RefI}, the energy current density $\tilde{j}_{\mathrm{E}}$
was the only one introduced, but in addition, only one BC was considered
therein, namely, the wavefunction $\psi$ tends to zero at $x=\pm\infty$
for a strictly neutral particle present on the real line. We also
discuss this issue in the next section.

We find that $\varrho_{\mathrm{E}}$ and $j_{\mathrm{E}}$ satisfy
the equation $\hat{\mathrm{E}}\varrho_{\mathrm{E}}-\hat{\mathrm{p}}j_{\mathrm{E}}=(\hat{\mathrm{E}}S)\psi^{*}\psi$,
which becomes a continuity equation only when $S$ does not explicitly
depend on time (or when it is zero). If we integrate this continuity
equation in the interval $\Omega=[a,b]$ and use the result given
in Eq. (27), we obtain the result $\int_{\Omega}\mathrm{d}x\,\varrho_{\mathrm{E}}=\mathrm{const}$.
In other words, if $\dot{S}=0$, then the pseudo self-adjointness
of $\hat{\mathrm{h}}$ implies that the result $\langle\langle\Psi,\hat{\mathrm{E}}\Psi\rangle\rangle=\mathrm{const}$
is satisfied for any solution of the equations of motion (see Eq.
(24)). Moreover, $\varrho_{\mathrm{E}}$ and $j_{\mathrm{E}}$ are
two components of a Lorentz second-rank tensor, an uncommon energy-momentum
tensor (see Appendix D). 

Similarly, $T_{\;\;0}^{0}$ and $cT_{\;\;0}^{1}\equiv\tilde{j}_{\mathrm{E}}$
satisfy the equation $\hat{\mathrm{E}}T_{\;\;0}^{0}-c\hat{\mathrm{p}}T_{\;\;0}^{1}=(\hat{\mathrm{E}}S)\psi^{*}\psi$,
that is, $\partial_{\mu}T_{\;\;0}^{\mu}=(\partial_{0}S)\psi^{*}\psi$.
From the latter result, if $\dot{S}=0$, then the result $\int_{\Omega}\mathrm{d}x\, T_{\;\;0}^{0}=\mathrm{const}$
is obtained only when Eq. (35) is verified (i.e., only when the first
``surface term'' in Eq. (34) vanishes). It is precisely the result
$\int_{\Omega}\mathrm{d}x\, T_{\;\;0}^{0}=\mathrm{const}$ that led
to the components $T_{\;\;0}^{0}$ and $cT_{\;\;0}^{1}$ being usually
considered the appropriate local observables associated with the energy
when $\psi$ describes a complex particle or a strictly neutral particle;
however, the result strongly depends on the BCs imposed. Then, the
pair $\{T_{\;\;0}^{0},cT_{\;\;0}^{1}\}$ (or $\{T_{\;\;0}^{0},\tilde{j}_{\mathrm{E}}\}$)
does not necessarily lead to a conserved quantity. 

As a matter of fact, we have the relation $\hat{\mathrm{E}}\varrho_{\mathrm{E}}-\hat{\mathrm{p}}j_{\mathrm{E}}=\hat{\mathrm{E}}T_{\;\;0}^{0}-\hat{\mathrm{p}}\tilde{j}_{\mathrm{E}}=\mathrm{i}\hbar c\,\partial_{\mu}T_{\;\;0}^{\mu}$;
thus, if $\dot{S}=0$, then $\hat{\mathrm{E}}\varrho_{\mathrm{E}}-\hat{\mathrm{p}}j_{\mathrm{E}}=0$
implies that $\hat{\mathrm{E}}T_{\;\;0}^{0}-c\hat{\mathrm{p}}T_{\;\;0}^{1}=0$,
and vice versa. Furthermore, because $\mathrm{i}\hbar c\,\partial_{\mu}T_{\;\;0}^{\mu}$
is a Lorentz scalar, $\hat{\mathrm{E}}\varrho_{\mathrm{E}}-\hat{\mathrm{p}}j_{\mathrm{E}}$
is also a Lorentz scalar, i.e., the continuity equation with the pair
$\{\varrho_{\mathrm{E}},j_{\mathrm{E}}\}$ can also be Lorentz covariant
(see Appendix D). 

Additionally, $\partial_{\mu}T_{\;\;1}^{\mu}=(\partial_{1}S)\psi^{*}\psi$,
and therefore, $\partial_{\mu}T_{\;\;\nu}^{\mu}=(\partial_{\nu}S)\psi^{*}\psi$
(see Appendix C). Certainly, $T_{\;\;1}^{0}$, a momentum density,
and $T_{\;\;1}^{1}$, a momentum current density, are the other two
components of the energy-momentum tensor $T_{\;\;\nu}^{\mu}$ (also
$T^{01}=T^{10}$, as expected for a scalar field) \cite{RefP}. On
the other hand, it can be shown that the pair $\{\tilde{\varrho}_{\mathrm{E}},cT_{\;\;0}^{1}\}$
(or $\{\tilde{\varrho}_{\mathrm{E}},\tilde{j}_{\mathrm{E}}\}$) does
not satisfy an equation that can lead to a continuity equation only
when $\dot{S}=0$.

In summary, if $\hat{\mathrm{h}}=\hat{\mathrm{h}}_{\mathrm{adj}}$
and if $\Psi$ and $\Phi$ satisfy the 1D FV equation, then $\langle\langle\Psi,\Phi\rangle\rangle=\mathrm{const}$
and therefore $\langle\langle\Psi,\Psi\rangle\rangle=\int_{\Omega}\mathrm{d}x\,\varrho$
is also a constant. The latter quantity is real, but it is not always
a positive number. On the other hand, if $\hat{\mathrm{h}}=\hat{\mathrm{h}}_{\mathrm{adj}}$
with $\dot{S}=0$ and if $\Psi$ and $\Phi$ satisfy the 1D FV equation,
then $\langle\langle\Psi,\hat{\mathrm{E}}\Phi\rangle\rangle=\mathrm{const}$,
and therefore $\langle\langle\Psi,\hat{\mathrm{E}}\Psi\rangle\rangle=\int_{\Omega}\mathrm{d}x\,\varrho_{\mathrm{E}}$
is also a constant. The latter quantity, the expectation value of
the energy operator in the state $\Psi$, is also a real number, and
because we are on-shell, it is the mean value of the Hamiltonian operator
in the same state, i.e., $\langle\langle\Psi,\hat{\mathrm{h}}\Psi\rangle\rangle$.
However, under certain realizable conditions, $\langle\langle\Psi,\hat{\mathrm{E}}\Psi\rangle\rangle$
can even be a positive quantity. In fact, using the Hamiltonian operator
in Eq. (5), we can write the following expression: 
\begin{equation}
\langle\langle\Psi,\hat{\mathrm{E}}\Psi\rangle\rangle=\langle\langle\Psi,\hat{\mathrm{h}}\Psi\rangle\rangle=\langle\langle\Psi,\hat{\mathrm{T}}\Psi\rangle\rangle+\langle\langle\Psi,\mathrm{m}c^{2}\hat{\tau}_{3}\Psi\rangle\rangle+\langle\langle\Psi,S(\hat{\tau}_{3}+\mathrm{i}\hat{\tau}_{2})\Psi\rangle\rangle,
\end{equation}
where $\hat{\mathrm{T}}$ is a kind of kinetic energy operator, namely,

\begin{equation}
\hat{\mathrm{T}}=\frac{\hat{\mathrm{p}}^{2}}{2\mathrm{m}}(\hat{\tau}_{3}+\mathrm{i}\hat{\tau}_{2}).
\end{equation}
We obtain the following result: 
\[
\langle\langle\Psi,\hat{\mathrm{E}}\Psi\rangle\rangle=\frac{\hbar}{2\mathrm{m}c}\left.\mathrm{Im}[\,\psi^{*}(c\hat{\mathrm{p}}\psi)\,]\right|_{a}^{b}+\frac{1}{2\mathrm{m}c^{2}}\int_{\Omega}\mathrm{d}x\left|\, c\hat{\mathrm{p}}\psi\,\right|^{2}+\frac{1}{2}\mathrm{m}c^{2}\int_{\Omega}\mathrm{d}x\left|\,\psi\,\right|^{2}
\]
\begin{equation}
+\frac{1}{2\mathrm{m}c^{2}}\int_{\Omega}\mathrm{d}x\mid\hat{\mathrm{E}}\psi\mid^{2}+\int_{\Omega}\mathrm{d}x\, S\left|\,\psi\,\right|^{2}.
\end{equation}
To obtain the latter result from Eq. (36), the integral in $\langle\langle\Psi,\hat{\mathrm{T}}\Psi\rangle\rangle$
was evaluated by parts once, and the boundary term that arose was
rewritten using $\psi^{*}\,\hat{\mathrm{p}}\psi=\mathrm{Re}(\psi^{*}\,\hat{\mathrm{p}}\psi)+\mathrm{i}\,\mathrm{Im}(\psi^{*}\,\hat{\mathrm{p}}\psi)$
and the result given in Eq. (16). The integral in $\langle\langle\Psi,\mathrm{m}c^{2}\hat{\tau}_{3}\Psi\rangle\rangle$
was expressed in terms of $\psi$, making use of the results arising
from the definition of $\Psi$ given in Eq. (13). Thus, $\langle\langle\Psi,\hat{\mathrm{E}}\Psi\rangle\rangle$
in Eq. (38) is always a real quantity; however, when the boundary
term vanishes (again, which depends on the BCs), $\langle\langle\Psi,\hat{\mathrm{E}}\Psi\rangle\rangle$
is definitely a positive quantity (but additionally, $S$ must be
a positive function in the interval $\Omega$). Certainly, the result
in Eq. (38) can be obtained immediately by integrating the energy
density given in Eq. (30) within the limits of $\Omega$, namely,
\begin{equation}
\langle\langle\Psi,\hat{\mathrm{E}}\Psi\rangle\rangle=\int_{\Omega}\mathrm{d}x\,\varrho_{\mathrm{E}}=\frac{\hbar}{2\mathrm{m}c}\left.\mathrm{Im}[\,\psi^{*}(c\hat{\mathrm{p}}\psi)\,]\right|_{a}^{b}+\int_{\Omega}\mathrm{d}x\, T_{\;\;0}^{0}
\end{equation}
(we recall that $\int_{\Omega}\mathrm{d}x\,\varrho=\mathrm{const}$;
hence, $\int_{\Omega}\mathrm{d}x\,\hat{\mathrm{E}}\varrho=0$). Thus,
$\varrho_{\mathrm{E}}$ is a complex local observable (when $\psi\in\mathbb{C}$),
but its integral in $\Omega$ is always a real quantity, which makes
it a physically realistic quantity. In general, when the scalar field
is studied, the term evaluated at the boundaries in Eq. (39) is neglected.
This neglect is argued for on the basis of the imposition of the Dirichlet
BC at $x=a$ and $x=b$ (or at $x=-\infty$ and $x=+\infty$ when
$\Omega$ is the whole real line). 

All the results presented above are valid for a 1D KFG particle. In
the next section, we address the conditions that a 1D KFGM particle
must satisfy. For example, we find the mathematical conditions that
ensure that the term evaluated at the boundaries in Eq. (38) or (39)
(and in Eq. (34)) vanishes. The latter occurs only for four confining
BCs and a one-parameter set of nonconfining BCs. We also give other
examples of BCs that do not meet these conditions. 

\section{Imposing the Majorana conditions}

\noindent In the characterization of a 1D KFGM particle, one has two
specific Majorana conditions, namely, a standard one, $\Psi=\Psi_{c}=\hat{\tau}_{1}\Psi^{*}$
($\Rightarrow\psi=\psi^{*}$), and a nonstandard one, $\Psi=-\Psi_{c}=-\hat{\tau}_{1}\Psi^{*}$
($\Rightarrow\psi=-\psi^{*}$) \cite{RefE}, where $\hat{\tau}_{1}=\hat{\sigma}_{x}$
is a Pauli matrix. When the solutions of the 1D KFG equation verify
the relations $\psi=\psi^{*}$ or $\psi=-\psi^{*}$ ($\Rightarrow\psi=\mathrm{i}\tilde{\psi}$,
where $\tilde{\psi}$ is a real wavefunction), the energy density
$\varrho_{\mathrm{E}}$ and the energy current density $j_{\mathrm{E}}$
become real quantities, and even better, they do not automatically
vanish (see Eqs. (25) and (26)); in addition, $j_{\mathrm{E}}$ satisfies
the condition given in Eq. (27). Likewise, $\tilde{\varrho}_{\mathrm{E}}$
and $T_{\;\;0}^{0}$, and $\tilde{j}_{\mathrm{E}}$, do not automatically
vanish either (see Eqs. (29), (31) and (33)), but $\tilde{j}_{\mathrm{E}}$
does not necessarily satisfy the condition given in Eq. (35). As will
be shown throughout this section, for a 1D KFGM particle, the energy
current density $j_{\mathrm{E}}$ considers impenetrable all the BCs
that were already considered impenetrable for a 1D KFG particle. Similarly,
it characterizes as penetrable those BCs that were already considered
penetrable, as would be expected. Likewise, the energy current density
$\tilde{j}_{\mathrm{E}}$ only succeeds in considering impenetrable
(penetrable) some of the BCs that are indeed impenetrable (penetrable).
Thus, this energy current density can provide only an incomplete description
of a 1D KFGM particle that is inside a finite interval. 

If we impose any of the two Majorana conditions on the four-parameter
general set of BCs for the solutions $\Psi$ of the 1D FV equation,
then it becomes a three-parameter general set, namely, 
\begin{equation}
\left[\begin{array}{c}
(\hat{\tau}_{3}+\mathrm{i}\hat{\tau}_{2})(\Psi-\mathrm{i}\lambda\Psi_{x})(b,t)\\
(\hat{\tau}_{3}+\mathrm{i}\hat{\tau}_{2})(\Psi+\mathrm{i}\lambda\Psi_{x})(a,t)
\end{array}\right]=\hat{\mathrm{U}}_{(4\times4)}\left[\begin{array}{c}
(\hat{\tau}_{3}+\mathrm{i}\hat{\tau}_{2})(\Psi+\mathrm{i}\lambda\Psi_{x})(b,t)\\
(\hat{\tau}_{3}+\mathrm{i}\hat{\tau}_{2})(\Psi-\mathrm{i}\lambda\Psi_{x})(a,t)
\end{array}\right],
\end{equation}
where $\lambda\in\mathbb{R}$, and $\hat{\mathrm{U}}_{(4\times4)}$
is the following $4\times4$ unitary and (complex) symmetric matrix:
\begin{equation}
\hat{\mathrm{U}}_{(4\times4)}=\mathrm{e}^{\mathrm{i}\,\mu}\left[\begin{array}{cc}
(\mathrm{m}_{0}-\mathrm{i}\,\mathrm{m}_{3})\hat{1}_{2} & -\mathrm{i}\,\mathrm{m}_{1}\hat{1}_{2}\\
-\mathrm{i}\,\mathrm{m}_{1}\hat{1}_{2} & (\mathrm{m}_{0}+\mathrm{i}\,\mathrm{m}_{3})\hat{1}_{2}
\end{array}\right],
\end{equation}
where $\mu\in[0,\pi)$, and the real quantities $\mathrm{m}_{0}$,
$\mathrm{m}_{1}$ and $\mathrm{m}_{3}$ satisfy $(\mathrm{m}_{0})^{2}+(\mathrm{m}_{1})^{2}+(\mathrm{m}_{3})^{2}=1$
(For clarity, we use the same notation used in Ref. \cite{RefE}.)
Similarly, the three-parameter general set of BCs for the (ultimately
real) solutions $\psi$ of the 1D KFG equation is given by 
\begin{equation}
\left[\begin{array}{c}
\psi(b,t)-\mathrm{i}\lambda\psi_{x}(b,t)\\
\psi(a,t)+\mathrm{i}\lambda\psi_{x}(a,t)
\end{array}\right]=\hat{\mathrm{U}}_{(2\times2)}\left[\begin{array}{c}
\psi(b,t)+\mathrm{i}\lambda\psi_{x}(b,t)\\
\psi(a,t)-\mathrm{i}\lambda\psi_{x}(a,t)
\end{array}\right],
\end{equation}
where the $2\times2$ unitary and (complex) symmetric matrix $\hat{\mathrm{U}}_{(2\times2)}$
is given by 
\begin{equation}
\hat{\mathrm{U}}_{(2\times2)}=\mathrm{e}^{\mathrm{i}\,\mu}\left[\begin{array}{cc}
\mathrm{m}_{0}-\mathrm{i}\,\mathrm{m}_{3} & -\mathrm{i}\,\mathrm{m}_{1}\\
-\mathrm{i}\,\mathrm{m}_{1} & \mathrm{m}_{0}+\mathrm{i}\,\mathrm{m}_{3}
\end{array}\right]
\end{equation}
(see Ref. \cite{RefE}). The general set of BCs given in Eq. (42)
is obtained by setting $\mathrm{m}_{2}=0$ in the most general set
of BCs given in Eq. (17). Imposing the Majorana conditions in (17)
leads to the result $\mathrm{m}_{2}=0$. The general set of BCs can
also be written as follows:
\begin{equation}
\left[\begin{array}{c}
(\hat{\mathrm{E}}\psi)(b,t)-\mathrm{i}\lambda(\hat{\mathrm{E}}\psi_{x})(b,t)\\
(\hat{\mathrm{E}}\psi)(a,t)+\mathrm{i}\lambda(\hat{\mathrm{E}}\psi_{x})(a,t)
\end{array}\right]=\hat{\mathrm{U}}_{(2\times2)}\left[\begin{array}{c}
(\hat{\mathrm{E}}\psi)(b,t)+\mathrm{i}\lambda(\hat{\mathrm{E}}\psi_{x})(b,t)\\
(\hat{\mathrm{E}}\psi)(a,t)-\mathrm{i}\lambda(\hat{\mathrm{E}}\psi_{x})(a,t)
\end{array}\right].
\end{equation}
The latter set is obtained by simply applying the energy operator
to both sides of the set given in Eq. (42). 

We can write an expression that gives us the value of $j_{\mathrm{E}}$
given in Eq. (26) at the walls of the box (which will be similar to
that already reported in Eq. (19) for the usual current $j$). Indeed,
this expression depends on the parameters that characterize the matrix
$\hat{\mathrm{U}}_{(2\times2)}$, in addition to the quantities $\psi$
and $\hat{\mathrm{E}}\psi$ evaluated at the extremes of the interval.
To obtain the result, the first step is to write the general set of
BCs in Eq. (42) as follows:
\begin{equation}
\left[\begin{array}{c}
\psi(b,t)\\
\lambda\psi_{x}(b,t)
\end{array}\right]=\hat{\mathrm{M}}\left[\begin{array}{c}
\psi(a,t)\\
\lambda\psi_{x}(a,t)
\end{array}\right],
\end{equation}
where the $2\times2$ matrix $\hat{\mathrm{M}}$ is given by
\begin{equation}
\hat{\mathrm{M}}=\frac{1}{\mathrm{m}_{1}}\left[\begin{array}{cc}
\mathrm{m}_{3}+\sin(\mu) & -\mathrm{m}_{0}-\cos(\mu)\\
-\mathrm{m}_{0}+\cos(\mu) & -\mathrm{m}_{3}+\sin(\mu)
\end{array}\right].
\end{equation}

\noindent Clearly, $\hat{\mathrm{M}}$ belongs to the special linear
group $\mathrm{SL}(2,\mathbb{R})$. In addition, the inverse expression
of Eq. (45) can be written immediately, and the respective inverse
matrix of $\hat{\mathrm{M}}$ can be obtained from the latter by making
the substitutions $\mathrm{m}_{0}\rightarrow-\mathrm{m}_{0}$, $\mathrm{m}_{3}\rightarrow-\mathrm{m}_{3}$
and $\cos(\mu)\rightarrow-\cos(\mu)$. Apparently, we do not have
within the family of BCs given in Eq. (45) the BCs for which $\mathrm{m}_{1}=0$;
however, we can obtain them if we first place the $\mathrm{m}_{1}$
term in Eq. (46) on the left-hand side of Eq. (45) and then make $\mathrm{m}_{1}=0$.
From this procedure, we obtain two relationships linking $\psi(a,t)$
with $\lambda\psi_{x}(a,t)$: namely,
\begin{equation}
\left(\mathrm{m}_{3}+\sin(\mu)\right)\psi(a,t)-\left(\mathrm{m}_{0}+\cos(\mu)\right)\lambda\psi_{x}(a,t)=0
\end{equation}
and 
\begin{equation}
\left(\mathrm{m}_{0}-\cos(\mu)\right)\psi(a,t)+\left(\mathrm{m}_{3}-\sin(\mu)\right)\lambda\psi_{x}(a,t)=0.
\end{equation}
Similarly, if we repeat the procedure with the inverse expression
of Eq. (45), we obtain the following two relations linking $\psi(b,t)$
with $\lambda\psi_{x}(b,t)$: 
\begin{equation}
\left(\mathrm{m}_{3}-\sin(\mu)\right)\psi(b,t)-\left(\mathrm{m}_{0}+\cos(\mu)\right)\lambda\psi_{x}(b,t)=0
\end{equation}
and 
\begin{equation}
\left(\mathrm{m}_{0}-\cos(\mu)\right)\psi(b,t)+\left(\mathrm{m}_{3}+\sin(\mu)\right)\lambda\psi_{x}(b,t)=0
\end{equation}
(recall that $\mu\in[0,\pi)$, and $(\mathrm{m}_{0})^{2}+(\mathrm{m}_{3})^{2}=1$
). The latter four relations generate all the BCs for which $\mathrm{m}_{1}=0$.
These $2+2=4$ relations can be reduced to only $1+1=2$ relations.
In fact, the latter can be obtained via $\mathrm{m}_{1}=0$ in Eq.
(42), i.e., in the matrix $\hat{\mathrm{U}}_{(2\times2)}$ in Eq.
(43). Each relation obtained from Eq. (42) leads to two relations
of the type mentioned just before. 

We can also write the time derivative of the general subfamily of
BCs given in Eq. (45), namely, 
\begin{equation}
\left[\begin{array}{c}
(\hat{\mathrm{E}}\psi)(b,t)\\
\lambda(\hat{\mathrm{E}}\psi_{x})(b,t)
\end{array}\right]=\hat{\mathrm{M}}\left[\begin{array}{c}
(\hat{\mathrm{E}}\psi)(a,t)\\
\lambda(\hat{\mathrm{E}}\psi_{x})(a,t)
\end{array}\right].
\end{equation}
From this matrix equation, we first write $\lambda(\hat{\mathrm{E}}\psi_{x})(a,t)$
in terms of $(\hat{\mathrm{E}}\psi)(a,t)$ and $(\hat{\mathrm{E}}\psi)(b,t)$.
Then, from the complex conjugate of the matrix equation given in Eq.
(45), we write $\lambda\psi_{x}^{*}(a,t)$ in terms of $\psi^{*}(a,t)$
and $\psi^{*}(b,t)$. Finally, by evaluating the energy current density
given in Eq. (26) at $x=a$ and using the results just obtained, we
obtain the following result:
\begin{equation}
j_{\mathrm{E}}(x=a,t)=\frac{\mathrm{i}\hbar}{2\mathrm{m}}\frac{1}{\lambda}\left(\frac{\mathrm{m}_{1}}{\mathrm{m}_{0}+\cos(\mu)}\right)\left[\,\psi^{*}(a,t)\,(\hat{\mathrm{E}}\psi)(b,t)-\psi^{*}(b,t)\,(\hat{\mathrm{E}}\psi)(a,t)\,\right],
\end{equation}
which, by virtue of the result given in Eq. (27), is also equal to
$j_{\mathrm{E}}(x=b,t)$. Clearly, when we set $\mathrm{m}_{1}=0$
in Eq. (52), then $j_{\mathrm{E}}(a,t)=j_{\mathrm{E}}(b,t)=0$. The
latter condition defines the confining BCs when the wavefunction is
real or purely imaginary. Thus, all confining BCs can be obtained
from Eq. (42) after making $\mathrm{m}_{1}=0$ or from Eq. (45) and
its inverse, i.e., from Eqs. (47)-(50). 

After $\mathrm{m}_{2}=0$, we make explicit use of the relation $\psi=\psi^{*}$,
or $\psi=-\psi^{*}$ (a Majorana condition). We automatically obtain
the following results: First, $j(a,t)=j(b,t)=0$, as expected (see
Eq. (19) and the discussion that follows this equation). Similarly,
it follows that $(\hat{\mathrm{E}}\, j)(a,t)=(\hat{\mathrm{E}}\, j)(b,t)=0$.
On the other hand, again making explicit use of a Majorana condition,
we still have that $j_{\mathrm{E}}(a,t)=j_{\mathrm{E}}(b,t)$, but
this equality is not automatically equal to zero, as required (see
Eq. (52)). We recall that, in this case, $j_{\mathrm{E}}$ is also
a real-valued quantity (see Eq. (26)). Finally, as a consequence of
the latter results, the relation given in Eq. (34), with $\psi=\psi^{*}$,
or $\psi=-\psi^{*}$, is automatically satisfied. 

Indeed, if the wavefunction is complex, $j_{\mathrm{E}}$ has the
disadvantage of being a complex quantity. This characteristic would
not be a problem as long as the spatial integral of $j_{\mathrm{E}}$
is a real quantity. In fact, it is easy to show that $\int_{\Omega}\mathrm{d}x\, j_{\mathrm{E}}$
is real only when $\int_{\Omega}\mathrm{d}x\, j=\mathrm{const}$ (see
Eq. (32)). As discussed above, the energy density $\varrho_{\mathrm{E}}$
is a complex quantity when $\psi\in\mathbb{C}$, but its spatial integral,
which is the mean value of $\hat{\mathrm{E}}$, is always real. Incidentally,
the mean value of the operator $c\hat{\mathrm{p}}$ is related to
the spatial integral of $j_{\mathrm{E}}$, as follows:
\begin{equation}
\langle\langle\Psi,c\hat{\mathrm{p}}\Psi\rangle\rangle=\frac{1}{c}\int_{\Omega}\mathrm{d}x\, j_{\mathrm{E}}-\frac{\mathrm{i}}{\mathrm{m}c^{2}}\int_{\Omega}\mathrm{d}x\,\mathrm{Im}[\,(\hat{\mathrm{E}}\psi^{*})\,(c\hat{\mathrm{p}}\psi)\,]
\end{equation}
(the latter result emerges after the definition of the scalar product
given in Eq. (14) and the definition of the energy current density
$j_{\mathrm{E}}$ in Eq. (26) are used). Note that the condition $\int_{\Omega}\mathrm{d}x\, j=\mathrm{const}$
is sufficient for $\int_{\Omega}\mathrm{d}x\, j_{\mathrm{E}}$ to
be real but not for $\langle\langle\Psi,c\hat{\mathrm{p}}\Psi\rangle\rangle$
to be real as well (see Eq. (53)). However, when $\psi$ satisfies
a Majorana condition, one has that
\begin{equation}
\langle\langle\Psi,c\hat{\mathrm{p}}\Psi\rangle\rangle=\frac{1}{c}\int_{\Omega}\mathrm{d}x\, j_{\mathrm{E}},
\end{equation}
and therefore, the mean value of $c\hat{\mathrm{p}}$ also becomes
a real quantity (because, in this case, $j_{\mathrm{E}}$ is real). 

At this point, certain remarks are in order. The general set of BCs
that is compatible with the relation $\left.\left[\, j(x,t)\,\right]\right|_{a}^{b}=0$
is given in Eqs. (17) and (18), but the solutions $\psi$ must be
complex (although not imaginary). If these solutions satisfy a Majorana
condition, then one has that $\left.\left[\, j(x,t)\,\right]\right|_{a}^{b}$
is trivially zero (because $j(x,t)=0$). The general set of BCs given
in Eqs. (17) and (18) is also compatible with the relation $\left.\left[\, j_{\mathrm{E}}(x,t)\,\right]\right|_{a}^{b}=0$,
and the solutions $\psi$ can be real or complex. If these solutions
satisfy a Majorana condition, then $j_{\mathrm{E}}(x,t)$ becomes
a real quantity and $\mathrm{m}_{2}=0$. Consequently, the general
set of BCs in Eqs. (17) and (18) becomes Eqs. (42) and (43) (or equivalently,
Eqs. (45) and (46) together with the inverse expressions of Eqs. (45)
and (46)). Similarly, to obtain the BCs compatible with the relation
$\left.\left[\,\tilde{j}_{\mathrm{E}}(x,t)\,\right]\right|_{a}^{b}=0$,
the conditions compatible with the relation $\left.\left[\,\mathrm{Im}(\psi\,(c\hat{\mathrm{p}}\psi))\,\right]\right|_{a}^{b}=0$
(see Eq. (34)) should first be obtained; if the latter relation is
verified, then $\left.\hat{\mathrm{E}}\left[\,\mathrm{Im}(\psi\,(c\hat{\mathrm{p}}\psi))\,\right]\right|_{a}^{b}=0$.
In addition, if $\left.\left[\,\mathrm{Im}(\psi\,(c\hat{\mathrm{p}}\psi))\,\right]\right|_{a}^{b}=0$,
then the mean value of the operator $\hat{\mathrm{E}}$ is necessarily
a positive quantity, provided that $S\geq0$ in the interval $\Omega$,
and both $\varrho_{\mathrm{E}}$ and $T_{\;\;0}^{0}$, when integrated
in $\Omega$, provide the correct expectation value $\langle\langle\Psi,\hat{\mathrm{E}}\Psi\rangle\rangle$
(see Eq. (39)). 

Thus, we start by writing the following algebraic expression: 
\begin{equation}
\mathrm{Im}(\psi\,(c\hat{\mathrm{p}}\psi))=-\hbar c\frac{1}{\lambda}\psi(x,t)\,\lambda\psi_{x}(x,t)=-\frac{\hbar c}{2}\frac{1}{\lambda}\left[\begin{array}{c}
\psi(x,t)\\
\lambda\psi_{x}(x,t)
\end{array}\right]^{\mathrm{T}}\hat{\tau}_{1}\left[\begin{array}{c}
\psi(x,t)\\
\lambda\psi_{x}(x,t)
\end{array}\right],
\end{equation}
where $\hat{\tau}_{1}=\hat{\sigma}_{x}$ is a Pauli matrix. If the
general set of BCs given in Eq. (45) is used, then the mathematical
condition that ensures the equality
\begin{equation}
\left.\left[\,\mathrm{Im}(\psi\,(c\hat{\mathrm{p}}\psi))\,\right]\right|_{a}^{b}=\left(\,\mathrm{Im}(\psi\,(c\hat{\mathrm{p}}\psi))\,\right)(b,t)-\left(\,\mathrm{Im}(\psi\,(c\hat{\mathrm{p}}\psi))\,\right)(a,t)=0
\end{equation}
is given by
\begin{equation}
\hat{\mathrm{M}}^{\mathrm{T}}\,\hat{\tau}_{1}\,\hat{\mathrm{M}}=\hat{\tau}_{1}.
\end{equation}
If the inverse expression of the set of BCs given in Eq. (45) is used,
then the mathematical condition takes the following form:
\begin{equation}
(\hat{\mathrm{M}}^{-1})^{\mathrm{T}}\,\hat{\tau}_{1}\,\hat{\mathrm{M}}^{-1}=\hat{\tau}_{1}.
\end{equation}
We recall that the set of BCs given in Eq. (45) and its inverse expression
compose the complete family of BCs when $\mathrm{m}_{2}=0$.

Substituting the matrix $\hat{\mathrm{M}}$ given in Eq. (46) into
Eq. (57), the following relations are obtained:
\begin{equation}
\left(\mathrm{m}_{3}\pm\sin(\mu)\right)\left(\mathrm{m}_{0}\mp\cos(\mu)\right)=0,
\end{equation}
and 
\begin{equation}
\left(\mathrm{m}_{0}-\cos(\mu)\right)\left(\mathrm{m}_{0}+\cos(\mu)\right)=0,
\end{equation}
where $(\mathrm{m}_{0})^{2}+(\mathrm{m}_{1})^{2}+(\mathrm{m}_{3})^{2}=1$
and $\mu\in[0,\pi)$. Equation (60) can also be written as follows:
\begin{equation}
(\mathrm{m}_{3})^{2}-\sin^{2}(\mu)=-(\mathrm{m}_{1})^{2}.
\end{equation}
Likewise, substituting the matrix $\hat{\mathrm{M}}^{-1}$ into Eq.
(58), the following relations are obtained:
\begin{equation}
\left(\mathrm{m}_{3}\pm\sin(\mu)\right)\left(\mathrm{m}_{0}\pm\cos(\mu)\right)=0,
\end{equation}
and 
\begin{equation}
\left(\mathrm{m}_{0}-\cos(\mu)\right)\left(\mathrm{m}_{0}+\cos(\mu)\right)=0.
\end{equation}
Equation (63) is identical to Eq. (60); therefore, the former is also
equal to the relation given in Eq. (61). Thus, the relations in Eqs.
(59), (60) and (62) must be verified for the relation given in Eq.
(56) to be satisfied. In the confining case, i.e., $\mathrm{m}_{1}=0$
($\Rightarrow(\mathrm{m}_{0})^{2}+(\mathrm{m}_{3})^{2}=1$), Eq. (61)
takes the form 
\begin{equation}
\left(\mathrm{m}_{3}-\sin(\mu)\right)\left(\mathrm{m}_{3}+\sin(\mu)\right)=0.
\end{equation}
This relation together with Eqs. (59) and (62) must be satisfied by
the confining BCs that verify the result given in Eq. (56).

As mentioned before, the most general subfamily of confining BCs can
also be obtained by making $\mathrm{m}_{1}=0$ in the most general
family of BCs given in Eq. (42). If we perform this procedure, we
obtain two separate relations (one relation evaluated at $x=a$ and
the other at $x=b$) that can be written in a single expression as
follows:
\[
\left[\begin{array}{cc}
1-\mathrm{e}^{\mathrm{i}\,\mu}(\mathrm{m}_{0}-\mathrm{i}\,\mathrm{m}_{3})\; & \;-\mathrm{i}-\mathrm{i}\,\mathrm{e}^{\mathrm{i}\,\mu}(\mathrm{m}_{0}-\mathrm{i}\,\mathrm{m}_{3})\\
0\; & \;0
\end{array}\right]\left[\begin{array}{c}
\psi(b,t)\\
\lambda\psi_{x}(b,t)
\end{array}\right]
\]
\begin{equation}
=\left[\begin{array}{cc}
0\; & \;0\\
1-\mathrm{e}^{\mathrm{i}\,\mu}(\mathrm{m}_{0}+\mathrm{i}\,\mathrm{m}_{3})\; & \;\mathrm{i}+\mathrm{i}\,\mathrm{e}^{\mathrm{i}\,\mu}(\mathrm{m}_{0}+\mathrm{i}\,\mathrm{m}_{3})
\end{array}\right]\left[\begin{array}{c}
\psi(a,t)\\
\lambda\psi_{x}(a,t)
\end{array}\right].
\end{equation}
The latter relation is a two-parameter subfamily of BCs (recall that
$\mu\in[0,\pi)$ and $(\mathrm{m}_{0})^{2}+(\mathrm{m}_{3})^{2}=1$).
Certainly, the subfamily of BCs arising from Eq. (65) by making the
replacements $\psi\rightarrow\hat{\mathrm{E}}\psi$ and $\psi_{x}\rightarrow\hat{\mathrm{E}}\psi_{x}$,
is also verified. Furthermore, it can be easily confirmed that each
of the two independent relationships that make up the subfamily given
in Eq. (65) is a real-valued relation when $\psi$ satisfies a Majorana
condition, which is because the quantities 
\[
\frac{1-\mathrm{e}^{\mathrm{i}\,\mu}(\mathrm{m}_{0}-\mathrm{i}\,\mathrm{m}_{3})}{-\mathrm{i}-\mathrm{i}\,\mathrm{e}^{\mathrm{i}\,\mu}(\mathrm{m}_{0}-\mathrm{i}\,\mathrm{m}_{3})}\quad\mathrm{and}\quad\frac{1-\mathrm{e}^{\mathrm{i}\,\mu}(\mathrm{m}_{0}+\mathrm{i}\,\mathrm{m}_{3})}{\mathrm{i}+\mathrm{i}\,\mathrm{e}^{\mathrm{i}\,\mu}(\mathrm{m}_{0}+\mathrm{i}\,\mathrm{m}_{3})},
\]
are real, as expected. The two relations linking $\psi(a,t)$ to $\lambda\psi_{x}(a,t)$
($\psi(b,t)$ to $\lambda\psi_{x}(b,t)$) given in Eqs. (47) and (48)
(Eqs. (49) and (50)) are equivalent to the relation linking $\psi(a,t)$
to $\lambda\psi_{x}(a,t)$ ($\psi(b,t)$ to $\lambda\psi_{x}(b,t)$)
that arises from Eq. (65).

Certainly, we can explicitly check that all the confining BCs lead
immediately to the relation $j_{\mathrm{E}}(a,t)=j_{\mathrm{E}}(b,t)=0$.
For example, by evaluating the energy current density in Eq. (26)
at $x=a$ and substituting the quantities $\lambda\psi_{x}^{*}(a,t)$
and $\lambda(\hat{\mathrm{E}}\psi_{x})(a,t)$ obtained from the subfamily
of the confining BCs given in Eq. (65), we automatically obtain the
result $j_{\mathrm{E}}(a,t)=0$, which, by virtue of the result in
Eq. (27), is equal to $j_{\mathrm{E}}(b,t)=0$. The subfamily of the
confining BCs given in Eq. (65) and the subfamily given in Eq. (45)
plus Eq. (46), with $\mathrm{m}_{1}\neq0$, contain all BCs that are
in the three-parameter general set of BCs given in Eq. (42). 

As examples of impenetrable BCs for the 1D KFGM particle, we have
the following (in all these cases we have that $\mathrm{m}_{1}=0$): 
\begin{lyxlist}{00.00.0000}
\item [{(i)}] $\psi(a,t)=\psi(b,t)=0$ 
\end{lyxlist}
($\mathrm{m}_{0}=-1$, $\mathrm{m}_{3}=0$ and $\mu=0$). This BC
is the Dirichlet BC. We restrict ourselves to $\mu\in[0,\pi)$; thus,
the matrix $\hat{\mathrm{U}}_{(2\times2)}$ in Eq. (18) is a one-to-one
parameterization of the group $U(2)$. If we had chosen $\mu\in[0,\pi]$
from the beginning, this BC, for example, could also be obtained by
making $\mathrm{m}_{0}=+1$, $\mathrm{m}_{3}=0$ and $\mu=\pi$. 
\begin{lyxlist}{00.00.0000}
\item [{(ii)}] $\lambda\psi_{x}(a,t)=\lambda\psi_{x}(b,t)=0$ 
\end{lyxlist}
($\mathrm{m}_{0}=+1$, $\mathrm{m}_{3}=0$ and $\mu=0$). This BC
is the Neumann BC.
\begin{lyxlist}{00.00.0000}
\item [{(iii)}] $\psi(a,t)=\lambda\psi_{x}(b,t)=0$ 
\end{lyxlist}
($\mathrm{m}_{0}=0$, $\mathrm{m}_{3}=+1$ and $\mu=\pi/2$). This
BC is a mixed BC.
\begin{lyxlist}{00.00.0000}
\item [{(iv)}] $\lambda\psi_{x}(a,t)=\psi(b,t)=0$ 
\end{lyxlist}
($\mathrm{m}_{0}=0$, $\mathrm{m}_{3}=-1$ and $\mu=\pi/2$). This
BC is another mixed BC.

The BCs (i), (ii), (iii) and (iv) satisfy Eqs. (59) and (62) and Eq.
(64); hence, $\left.\left[\,\mathrm{Im}(\psi\,(c\hat{\mathrm{p}}\psi))\,\right]\right|_{a}^{b}=0$.
Furthermore, each term of the subtraction in Eq. (56) is also zero.
Thus, from the formula given in Eq. (34), it follows that $\left.\left[\,\tilde{j}_{\mathrm{E}}\,\right]\right|_{a}^{b}=0$.
Additionally, for these four confining BCs, the energy current density
$\tilde{j}_{\mathrm{E}}$ given in Eq. (33), i.e.,
\begin{equation}
\tilde{j}_{\mathrm{E}}=\frac{\mathrm{i}\hbar}{\mathrm{m}}\frac{1}{\lambda}(\hat{\mathrm{E}}\psi)(x,t)\,\lambda\psi_{x}(x,t)
\end{equation}
cancels out at the ends of the interval, i.e., $\tilde{j}_{\mathrm{E}}(b,t)=\tilde{j}_{\mathrm{E}}(a,t)=0$.
Then, using the Dirichlet BC $\psi(a=-\infty,t)=\psi(b=+\infty,t)=0$,
for a strictly neutral particle existing in the interval $\Omega=\mathbb{R}$,
the energy current density $j_{\mathrm{E}}$ vanishes at $x=\pm\infty$,
but in addition $\tilde{j}_{\mathrm{E}}$ also vanishes there. In
Ref. \cite{RefI}, only $\tilde{j}_{\mathrm{E}}$ was considered.
\begin{lyxlist}{00.00.0000}
\item [{(v)}] $\psi(a,t)-\lambda\psi_{x}(a,t)=0$ and $\psi(b,t)+\lambda\psi_{x}(b,t)=0$ 
\end{lyxlist}
($\mathrm{m}_{0}=+1$, $\mathrm{m}_{3}=0$ and $\mu=\pi/2$). This
BC is a type of Robin BC. 
\begin{lyxlist}{00.00.0000}
\item [{(vi)}] $\psi(a,t)+\lambda\psi_{x}(a,t)=0$ and $\psi(b,t)-\lambda\psi_{x}(b,t)=0$ 
\end{lyxlist}
($\mathrm{m}_{0}=-1$, $\mathrm{m}_{3}=0$ and $\mu=\pi/2$). This
BC is another Robin-type BC. 

The latter two confining BCs would be KFG versions of the BC commonly
used in the one-dimensional MIT bag model for hadronic structures
\cite{RefQ}. However, these two BCs do not satisfy Eqs. (59), (62)
and (64); hence, $\left.\left[\,\mathrm{Im}(\psi\,(c\hat{\mathrm{p}}\psi))\,\right]\right|_{a}^{b}\neq0$.
Consequently, from the formula given in Eq. (34), it follows that
$\left.\left[\,\tilde{j}_{\mathrm{E}}\,\right]\right|_{a}^{b}\neq0$.
In fact, for these two confining BCs, the energy current density $\tilde{j}_{\mathrm{E}}$
in Eq. (66) does not vanish at the walls of the interval, i.e., $\tilde{j}_{\mathrm{E}}(b,t)\neq0$
and $\tilde{j}_{\mathrm{E}}(a,t)\neq0$; however, we have that $j_{\mathrm{E}}(a,t)=j_{\mathrm{E}}(b,t)=0$.

The only confining BCs that satisfy the result given in Eq. (56) are
the BCs (i), (ii), (iii) and (iv). In fact, from the two relations
in Eq. (59), the relations $\mathrm{m}_{0}\,\mathrm{m}_{3}=\sin(\mu)\cos(\mu)$
and $\mathrm{m}_{0}\sin(\mu)=\mathrm{m}_{3}\cos(\mu)$ are obtained.
Similarly, from the two relations in Eq. (62), the relations $\mathrm{m}_{0}\,\mathrm{m}_{3}=-\sin(\mu)\cos(\mu)$
and $\mathrm{m}_{0}\sin(\mu)=-\mathrm{m}_{3}\cos(\mu)$ are obtained.
The latter two pairs of relations imply that $\mathrm{m}_{0}\,\mathrm{m}_{3}=0$,
$\sin(\mu)\cos(\mu)=0$, $\mathrm{m}_{0}\sin(\mu)=0$ and $\mathrm{m}_{3}\cos(\mu)=0$.
These four relations plus $(\mathrm{m}_{0})^{2}=\cos^{2}(\mu)$ (Eq.
(60)) and $(\mathrm{m}_{3})^{2}=\sin^{2}(\mu)$ (Eq. (64)) ($\Rightarrow(\mathrm{m}_{0})^{2}+(\mathrm{m}_{3})^{2}=1$)
give us only four solutions (or BCs), namely, $\mathrm{m}_{0}=0$,
$\mathrm{m}_{3}=\pm1$ and $\mu=\pi/2$ (these parameters generate
BCs (iii) and (iv)); and $\mathrm{m}_{0}=\pm1$, $\mathrm{m}_{3}=0$
and $\mu=0$ (these parameters generate BCs (i) and (ii)).

As examples of nonconfining BCs for the 1D KFGM particle, we have
the following (naturally, in all these cases, we have that $\mathrm{m}_{1}\neq0$): 
\begin{lyxlist}{00.00.0000}
\item [{(vii)}] $\psi(a,t)=\psi(b,t)$ and $\lambda\psi_{x}(a,t)=\lambda\psi_{x}(b,t)$ 
\end{lyxlist}
($\mathrm{m}_{0}=\mathrm{m}_{3}=0$, $\mathrm{m}_{1}=+1$ and $\mu=\pi/2$).
This BC is the periodic BC.
\begin{lyxlist}{00.00.0000}
\item [{(viii)}] $\psi(a,t)=-\psi(b,t)$ and $\lambda\psi_{x}(a,t)=-\lambda\psi_{x}(b,t)$ 
\end{lyxlist}
($\mathrm{m}_{0}=\mathrm{m}_{3}=0$, $\mathrm{m}_{1}=-1$ and $\mu=\pi/2$).
This BC is the antiperiodic BC. The following subfamily of nonconfining
BCs is a type of generalization of the latter two BCs:
\begin{lyxlist}{00.00.0000}
\item [{(ix)}] $\left[\begin{array}{c}
\psi(b,t)\\
\lambda\psi_{x}(b,t)
\end{array}\right]=\pm\left[\begin{array}{cc}
\sin(\mu) & -\cos(\mu)\\
\cos(\mu) & \sin(\mu)
\end{array}\right]\left[\begin{array}{c}
\psi(a,t)\\
\lambda\psi_{x}(a,t)
\end{array}\right]$ 
\[
=\pm\mathrm{e}^{-\mathrm{i}\,\hat{\sigma}_{y}(\frac{\pi}{2}-\mu)}\left[\begin{array}{c}
\psi(a,t)\\
\lambda\psi_{x}(a,t)
\end{array}\right]
\]

\end{lyxlist}
($\mathrm{m}_{0}=\mathrm{m}_{3}=0$, the upper sign corresponds to
$\mathrm{m}_{1}=+1$ and the lower sign to $\mathrm{m}_{1}=-1$, and
$\mu\in[0,\pi)$). By setting $\mu=\pi/2$ with $\mathrm{m}_{1}=+1$
in the latter set of BCs, one obtains the periodic BC, and setting
$\mathrm{m}_{1}=-1$, one obtains the antiperiodic BC. By choosing
$\mu=0$ in the latter subfamily, one obtains the following two BCs:
\begin{lyxlist}{00.00.0000}
\item [{(x)}] $\psi(a,t)=\pm\lambda\psi_{x}(b,t)$ and $\psi(b,t)=\mp\lambda\psi_{x}(a,t)$ 
\end{lyxlist}
($\mathrm{m}_{0}=\mathrm{m}_{3}=0$, the upper sign corresponds to
$\mathrm{m}_{1}=+1$ and the lower sign corresponds to $\mathrm{m}_{1}=-1$,
and $\mu=0$). These two BCs are a particular type of mixed BCs.

Of all the nonconfining BCs, only those that are within the following
one-parameter subfamily of BCs satisfy Eqs. (59), (60) and (62):
\begin{equation}
\left[\begin{array}{c}
\psi(b,t)\\
\lambda\psi_{x}(b,t)
\end{array}\right]=\frac{1}{\mathrm{m}_{1}}\left[\begin{array}{cc}
\mathrm{m}_{3}+1 & 0\\
0 & -\mathrm{m}_{3}+1
\end{array}\right]\left[\begin{array}{c}
\psi(a,t)\\
\lambda\psi_{x}(a,t)
\end{array}\right],
\end{equation}
where $(\mathrm{m}_{1})^{2}+(\mathrm{m}_{3})^{2}=1$. In fact, from
the two relations in Eq. (59) and the two relations in Eq. (62), the
relations $\mathrm{m}_{0}\,\mathrm{m}_{3}=0$, $\sin(\mu)\cos(\mu)=0$,
$\mathrm{m}_{0}\sin(\mu)=0$ and $\mathrm{m}_{3}\cos(\mu)=0$ are
obtained. The latter four relations plus $(\mathrm{m}_{0})^{2}=\cos^{2}(\mu)$
(Eq. (60)) give us only the following solutions: $\mathrm{m}_{0}=0$
and $\mu=\pi/2$. By substituting the latter parameters into Eq. (45),
one obtains the result given in Eq. (67). For all the BCs within the
subfamily in Eq. (67), we have that $\left.\left[\,\mathrm{Im}(\psi\,(c\hat{\mathrm{p}}\psi))\,\right]\right|_{a}^{b}=0$
(Eq. (56)), and therefore, $\left.\left[\,\tilde{j}_{\mathrm{E}}\,\right]\right|_{a}^{b}=0$
(Eq. (35)). Moreover, $\mathrm{Im}(\psi\,(c\hat{\mathrm{p}}\psi))$
and $\tilde{j}_{\mathrm{E}}$ do not cancel at the extremes of the
interval. Thus, for the remaining nonconfining BCs that are outside
the subfamily in Eq. (67), only the observable $j_{\mathrm{E}}$ can
adequately characterize the penetrability of the interval (because
only $j_{\mathrm{E}}$ satisfies the relation $\left.\left[\, j_{\mathrm{E}}(x,t)\,\right]\right|_{a}^{b}=0$).
Certainly, the periodic BC ((vii)) and the antiperiodic BC ((viii))
are within the subfamily in Eq. (67). 

As discussed above, a 1D KFGM particle supports only those BCs that
satisfy the condition $\mathrm{m}_{2}=0$. As examples of BCs that
are not suitable for a 1D KFGM particle but are suitable for a 1D
KFG particle (i.e., $\mathrm{m}_{2}\neq0$), we have the following: 
\begin{lyxlist}{00.00.0000}
\item [{(xi)}] $\psi(a,t)=\pm\mathrm{i}\psi(b,t)$ and $\lambda\psi_{x}(a,t)=\pm\mathrm{i}\lambda\psi_{x}(b,t)$ 
\end{lyxlist}
($\mathrm{m}_{0}=\mathrm{m}_{1}=\mathrm{m}_{3}=0$, the upper sign
corresponds to $\mathrm{m}_{2}=+1$ and the lower sign corresponds
to $\mathrm{m}_{2}=-1$, and $\mu=\pi/2$). These BCs are two types
of quasiperiodic and quasiantiperiodic BCs. 
\begin{lyxlist}{00.00.0000}
\item [{(xii)}] $\psi(a,t)=\pm\mathrm{i}\lambda\psi_{x}(b,t)$ and $\psi(b,t)=\pm\mathrm{i}\lambda\psi_{x}(a,t)$ 
\end{lyxlist}
($\mathrm{m}_{0}=\mathrm{m}_{1}=\mathrm{m}_{3}=0$, the upper sign
corresponds to $\mathrm{m}_{2}=+1$ and the lower sign corresponds
to $\mathrm{m}_{2}=-1$, and $\mu=0$). These BCs are two BCs of the
quasimixed type. Clearly, the BCs in (xi) and (xii) are complex BCs.
Incidentally, they are also nonconfining or permeable BCs (because
$\mathrm{m}_{2}\neq0$, and hence, $j(b,t)=j(a,t)$).

Finally, we discuss the problem of the energy currents that arise
when the energy current densities $j_{\mathrm{E}}$ and $\tilde{j}_{\mathrm{E}}$
are integrated in space. The relationship that connects these two
energy current densities is Eq. (32), namely, 
\begin{equation}
j_{\mathrm{E}}=\frac{\mathrm{i}}{2\mathrm{m}c}\, c\hat{\mathrm{p}}\left[\,\mathrm{Im}(\psi^{*}\,(\hat{\mathrm{E}}\psi))\,\right]+\frac{1}{2}\,\hat{\mathrm{E}}\, j+\tilde{j}_{\mathrm{E}}.
\end{equation}
To write this last relation, we also use the identity $\hat{\mathrm{E}}\left[\,\mathrm{Im}(\psi^{*}\,(c\hat{\mathrm{p}}\psi))\,\right]=c\hat{\mathrm{p}}\,[\,\mathrm{Im}(\psi^{*}\,(\hat{\mathrm{E}}\psi))\,]$.
Now, making explicit use of a Majorana condition, i.e., $\psi=\psi^{*}$,
or $\psi=-\psi^{*}$ ($\Rightarrow j=0$), Eq. (68) takes the following
form:
\begin{equation}
j_{\mathrm{E}}=\frac{\hbar}{2\mathrm{m}}\,\frac{\partial}{\partial x}\left[\,\mathrm{Im}(\psi\,(\hat{\mathrm{E}}\psi))\,\right]+\tilde{j}_{\mathrm{E}}.
\end{equation}
Integrating the latter relation over the finite interval $\Omega$,
we obtain the following result: 
\begin{equation}
J_{\mathrm{E}}=\frac{\hbar}{2\mathrm{m}}\left.\left[\,\mathrm{Im}(\psi\,(\hat{\mathrm{E}}\psi))\,\right]\right|_{a}^{b}+\tilde{J}_{\mathrm{E}},
\end{equation}
where $J_{\mathrm{E}}=\int_{\Omega}\mathrm{d}x\, j_{\mathrm{E}}$
and $\tilde{J}_{\mathrm{E}}=\int_{\Omega}\mathrm{d}x\,\tilde{j}_{\mathrm{E}}$
are two currents of energy. Clearly, if the particles were on the
entire real line ($\Omega=(-\infty,+\infty)$), and one assumes that
$\psi(a=-\infty,t)=\psi(b=+\infty,t)=0$, then the term evaluated
at the boundaries of $\Omega$ disappears and there would be no distinction
between $J_{\mathrm{E}}$ and $\tilde{J}_{\mathrm{E}}$. The latter
BC corresponds to the most common assumption of classical field theory,
i.e., \textquotedblleft{}the fields vanish at infinity.\textquotedblright{}
Certainly, this BC was the only BC used in Ref. \cite{RefI}. However,
we can obtain here the specific algebraic conditions under which the
BCs that are compatible with the result $\left.\left[\,\mathrm{Im}(\psi\,(\hat{\mathrm{E}}\psi))\,\right]\right|_{a}^{b}=0$
must obey.

Let us start by writing the following algebraic relation:
\begin{equation}
\mathrm{Im}(\psi\,(\hat{\mathrm{E}}\psi))=\frac{1}{\mathrm{i}}\psi(x,t)\,(\hat{\mathrm{E}}\psi)(x,t)=\frac{1}{2\mathrm{i}}\left[\begin{array}{c}
\psi(x,t)\\
\lambda\psi_{x}(x,t)
\end{array}\right]^{\mathrm{T}}\left(\hat{1}_{2}+\hat{\tau}_{3}\right)\left[\begin{array}{c}
(\hat{\mathrm{E}}\psi)(x,t)\\
\lambda(\hat{\mathrm{E}}\psi_{x})(x,t)
\end{array}\right].
\end{equation}
If we use the general family of BCs given in Eq. (45) (and Eq. (51)),
then the matrix condition that ensures the relation 
\begin{equation}
\left.\left[\,\mathrm{Im}(\psi\,(\hat{\mathrm{E}}\psi))\,\right]\right|_{a}^{b}=\left(\,\mathrm{Im}(\psi\,(\hat{\mathrm{E}}\psi))\,\right)(b,t)-\left(\,\mathrm{Im}(\psi\,(\hat{\mathrm{E}}\psi))\,\right)(a,t)=0
\end{equation}
is given by 
\begin{equation}
\hat{\mathrm{M}}^{\mathrm{T}}\left(\hat{1}_{2}+\hat{\tau}_{3}\right)\hat{\mathrm{M}}=\hat{1}_{2}+\hat{\tau}_{3}.
\end{equation}
If we use the inverse expression of Eqs. (45) and (51), then the matrix
condition is given by 
\begin{equation}
(\hat{\mathrm{M}}^{-1})^{\mathrm{T}}\left(\hat{1}_{2}+\hat{\tau}_{3}\right)\hat{\mathrm{M}}^{-1}=\hat{1}_{2}+\hat{\tau}_{3}.
\end{equation}

Substituting the matrix $\hat{\mathrm{M}}$ given in Eq. (46) into
Eq. (73), the following relations are obtained: 
\begin{equation}
\left(\mathrm{m}_{3}+\sin(\mu)\right)\left(\mathrm{m}_{0}+\cos(\mu)\right)=0,
\end{equation}
\begin{equation}
\left(\mathrm{m}_{3}+\sin(\mu)\right)^{2}=(\mathrm{m}_{1})^{2},
\end{equation}
and 
\begin{equation}
\left(\mathrm{m}_{0}+\cos(\mu)\right)^{2}=0.
\end{equation}
Similarly, substituting the matrix $\hat{\mathrm{M}}^{-1}$ into Eq.
(74), we obtain the following relations:
\begin{equation}
\left(-\mathrm{m}_{3}+\sin(\mu)\right)\left(\mathrm{m}_{0}+\cos(\mu)\right)=0,
\end{equation}
\begin{equation}
\left(-\mathrm{m}_{3}+\sin(\mu)\right)^{2}=(\mathrm{m}_{1})^{2},
\end{equation}
and 
\begin{equation}
\left(\mathrm{m}_{0}+\cos(\mu)\right)^{2}=0
\end{equation}
(this equation is Eq. (77)), where $(\mathrm{m}_{0})^{2}+(\mathrm{m}_{1})^{2}+(\mathrm{m}_{3})^{2}=1$
and $\mu\in[0,\pi)$. Thus, all the latter relations (Eqs. (75)-(79))
must be fulfilled for the condition given in Eq. (72) to be verified. 

It can be shown that the only confining BC that satisfies the result
$\left.\left[\,\mathrm{Im}(\psi\,(\hat{\mathrm{E}}\psi))\,\right]\right|_{a}^{b}=0$
(Eq. (72)) is the Dirichlet BC. Note that if $\mathrm{m}_{1}=0$,
Eqs. (76) and (79) lead to the results $\mathrm{m}_{3}=0$ and $\mu=0$.
Consequently, Eqs. (75) and (78) are automatically satisfied, and
Eq. (77) gives the result $\mathrm{m}_{0}=-1$. All these parameters
lead only to the Dirichlet BC. Thus, of the four confining BC that
satisfy the result $\left.\left[\,\mathrm{Im}(\psi\,(c\hat{\mathrm{p}}\psi))\,\right]\right|_{a}^{b}=0$
(Eq. (56)), only the Dirichlet BC ((i)) can also fulfill the result
$\left.\left[\,\mathrm{Im}(\psi\,(\hat{\mathrm{E}}\psi))\,\right]\right|_{a}^{b}=0$. 

Similarly, it can also be shown that the only nonconfining BCs that
satisfy the result $\left.\left[\,\mathrm{Im}(\psi\,(\hat{\mathrm{E}}\psi))\,\right]\right|_{a}^{b}=0$
are those included in the following one-parameter subfamily of BCs:
\begin{equation}
\left[\begin{array}{c}
\psi(b,t)\\
\lambda\psi_{x}(b,t)
\end{array}\right]=\pm\left[\begin{array}{cc}
1 & 0\\
2\cot(\mu) & 1
\end{array}\right]\left[\begin{array}{c}
\psi(a,t)\\
\lambda\psi_{x}(a,t)
\end{array}\right]
\end{equation}
where $\mu\in(0,\pi)$. Note that Eq. (77) gives the result $\mathrm{m}_{0}=-\cos(\mu)$.
Similarly, Eqs. (76) and (79) give the results $\mathrm{m}_{3}=0$
and $(\mathrm{m}_{1})^{2}=\sin^{2}(\mu)\,(\neq0)$. Consequently,
Eqs. (75) and (78) are automatically satisfied. All these parameters
lead only to the result in Eq. (81). Thus, of all the nonconfining
BCs that satisfy the result $\left.\left[\,\mathrm{Im}(\psi\,(c\hat{\mathrm{p}}\psi))\,\right]\right|_{a}^{b}=0$,
i.e., the BCs within the subfamily in Eq. (67), only the periodic
((vii)) and antiperiodic ((viii)) BCs can also fulfill the result
$\left.\left[\,\mathrm{Im}(\psi\,(\hat{\mathrm{E}}\psi))\,\right]\right|_{a}^{b}=0$.
Indeed, the subfamily in Eq. (67) requires that $\mu=\pi/2$; then,
if one imposes the latter parameter on the subfamily of BCs in Eq.
(81), only these two BCs emerge.

\section{Final discussion}

\noindent The usual density current $j=j(x,t)$ is useful for characterizing
a 1D KFG particle---a charged or ``complex particle''---but is not
useful for characterizing a 1D KFGM particle---a truly neutral or
``real particle''---because the real solutions of the KFG equation
cancel it out everywhere. However, the energy current density $j_{\mathrm{E}}=j_{\mathrm{E}}(x,t)$,
in general, does not serve to characterize a 1D KFG particle present
in an interval because, beyond being a complex quantity, its spatial
integral is a complex quantity unless the condition $\int_{\Omega}\mathrm{d}x\, j=\mathrm{const}$
is satisfied (see Eq. (32)). The latter is a difficult condition to
achieve if one considers solutions of the time-dependent 1D KFG equation.
However, $j_{\mathrm{E}}=j_{\mathrm{E}}(x,t)$ appears to be acceptable
for characterizing the confinement of a 1D KFGM particle that lies
within an interval. We recall that, in this case, this energy current
density is a real quantity, and its spatial integral is appropriately
a real quantity (see Eqs. (68) and (69)).

The complex solutions of the 1D KFG equation that describes a charged
particle confined to a finite interval satisfy an impenetrability
condition that can be expressed in terms of the usual current density,
namely, $j(a,t)=j(b,t)=0$. Among all the BCs that make up the four-parameter
general set of BCs for these solutions, only those that satisfy $\mathrm{m}_{1}=\mathrm{m}_{2}=0$
are impenetrable BCs. However, the solutions of the 1D KFG equation
describing a strictly neutral particle inside an interval satisfy
a Majorana condition ($\psi=\psi^{*}$ or $\psi=-\psi^{*}$), which
implies that all BCs included in the four-parameter general set of
BCs must satisfy the condition $\mathrm{m}_{2}=0$. If these particles
are confined to an interval, then they must also satisfy an impenetrability
condition, which can be expressed in terms of the energy current density
$j_{\mathrm{E}}=j_{\mathrm{E}}(x,t)$, namely, $j_{\mathrm{E}}(a,t)=j_{\mathrm{E}}(b,t)=0$.
Among all the BCs that are included in the three-parameter general
set of BCs, only those that satisfy $\mathrm{m}_{1}=0$ are confining.
These results indicate that if we can say that a BC is confining for
a strictly neutral particle, then we can also say that this BC is
confining for a charged particle and vice versa. In other words, the
BCs that confine a 1D KFGM particle also confine a 1D KFG particle
and vice versa. Similarly, any nonconfining BC for a 1D KFGM particle
is also nonconfining for a 1D KFG particle; however, the converse
is inaccurate because, for a 1D KFG particle, there are nonconfining
BCs that are unacceptable for a 1D KFGM particle. This unacceptability
emerges because all the BCs for a 1D KFGM particle must always verify
$\mathrm{m}_{2}=0$. Ultimately, more nonconfining BCs are available
for a 1D KFG particle than for a 1D KFGM particle. 

Among all the confining BCs that can be imposed on the solutions of
the 1D KFG equation describing a 1D KFGM particle, only four allow
us to ensure that the mean value $\langle\langle\Psi,\hat{\mathrm{E}}\Psi\rangle\rangle=\int_{\Omega}\mathrm{d}x\,\varrho_{\mathrm{E}}=\int_{\Omega}\mathrm{d}x\, T_{\;\;0}^{0}$
will undoubtedly be a positive quantity (with $S\geq0$). The Dirichlet
BC is one of these BCs, as expected. Furthermore, for these four confining
BCs only, the energy current density $\tilde{j}_{\mathrm{E}}=\tilde{j}_{\mathrm{E}}(x,t)$
is consistent with the impenetrable nature of the interval, i.e.,
$\tilde{j}_{\mathrm{E}}\equiv c\, T_{\;\;0}^{1}$ satisfies $\left.\left[\,\tilde{j}_{\mathrm{E}}\,\right]\right|_{a}^{b}=0$
(Eq. (35)) and the impenetrability condition $\tilde{j}_{\mathrm{E}}(a,t)=\tilde{j}_{\mathrm{E}}(b,t)=0$.
All the other confining BCs---an infinite number of confining BCs---satisfy
$\left.\left[\, j_{\mathrm{E}}\,\right]\right|_{a}^{b}=0$, but in
addition, $j_{\mathrm{E}}(a,t)=j_{\mathrm{E}}(b,t)=0$; however, they
satisfy neither $\left.\left[\,\tilde{j}_{\mathrm{E}}\,\right]\right|_{a}^{b}=0$
nor the impenetrability condition imposed on $\tilde{j}_{\mathrm{E}}$,
namely, $\tilde{j}_{\mathrm{E}}(a,t)=\tilde{j}_{\mathrm{E}}(b,t)=0$.
Certainly, the abovementioned four confining BCs satisfy the impenetrability
condition $j_{\mathrm{E}}(a,t)=j_{\mathrm{E}}(b,t)=0$. On the other
hand, only the BCs within a one-parameter subfamily of nonconfining
BCs also ensure that $\langle\langle\Psi,\hat{\mathrm{E}}\Psi\rangle\rangle$
is a positive quantity, and the energy current density $\tilde{j}_{\mathrm{E}}$
just satisfies the condition $\left.\left[\,\tilde{j}_{\mathrm{E}}\,\right]\right|_{a}^{b}=0$,
which is consistent with a penetrable interval. Incidentally, the
Dirichlet BC is the only impenetrable BC that ensures that the space
integrals of $j_{\mathrm{E}}$ and $\tilde{j}_{\mathrm{E}}$ are equal,
i.e., in this case, the respective energy currents are equal. Similarly,
only the BCs included in a one-parameter subfamily of nonconfining
BCs lead to this same result. 

As discussed before, the commonly used energy density $T_{\;\;0}^{0}$
and energy current density $\tilde{j}_{\mathrm{E}}\equiv c\, T_{\;\;0}^{1}$
do not necessarily lead to a globally conserved quantity when the
Lorentz scalar potential is time independent or equal to zero. Certainly,
this energy current density does not characterize all the BCs that
the system can have as impenetrable or penetrable; however, for some
specific BCs, it does. This strong dependence on BCs is typical of
quantum systems in which the particles are restricted to finite regions
in a few dimensions, i.e., low-dimensional systems. Thus, our work
is a pertinent generalization of the simplest problem of classical
field theory in (1+1) dimensions, which is the real scalar field problem
(with a Lorentz scalar potential), but this time in a finite interval.
Our results revealed a wide variety of acceptable BCs. The first choice
that leads to imposing the Dirichlet BC on the one-component wavefunction
$\psi$ greatly simplifies the study of the problem but sets aside
many models that are also realizable. Notably, the results presented
here can be used in the study of the Casimir effect problem for real
scalar fields in (1+1) dimensions \cite{RefR}. In fact, different
approaches to the study of this problem, both for real and complex
fields, with or without mass, and using various BCs presented here
have already been considered (see, for example, Refs. \cite{RefS,RefT,RefU,RefV,RefW}).

Thus, the densities $\varrho_{\mathrm{E}}$ and $j_{\mathrm{E}}$
allow a 1D KFGM particle confined or unconstrained to an interval
to be fully characterized and can also satisfy a continuity equation
that provides a global quantity that is constant over time. These
two quantities are two of the components of a second-rank Lorentz
tensor---an uncommon energy-momentum tensor--- $K_{\;\;\nu}^{\mu}$,
but in addition, this tensor is symmetric ($K^{01}=K^{10}$) only
when it is calculated for solutions of the 1D KFG equation that satisfy
a Majorana condition (see Appendix D). Here, we also identified the
BCs that lead to the same results that would be obtained if the components
of the standard energy-momentum tensor $T_{\;\;0}^{0}$ and $T_{\;\;0}^{1}=\tilde{j}_{\mathrm{E}}/c$
were used, i.e., in this case, we can say that the pairs of local
observables $\{T_{\;\;0}^{0},cT_{\;\;0}^{1}\}$ and $\{\varrho_{\mathrm{E}},j_{\mathrm{E}}\}=\{K_{\;\;0}^{0},cK_{\;\;0}^{1}\}$
are equally satisfactory (see Appendix D). As we have shown, this
occurs not only for the commonly used Dirichlet BC but also for three
other confining BCs and all BCs that are within a one-parameter family
of nonconfining BCs. Incidentally, reference \cite{RefE} also studied
1D KFGM particles that are within a finite interval; however, it did
not provide any way to differentiate between impermeable boundaries
and permeable boundaries. Our work is the first to address this specific
problem. 

Last, it is worth mentioning that the search for a physically acceptable
probability 4-current density (not a charge 4-current density) in
(3+1) dimensions, for charged and even strictly neutral scalar particles,
has been a long-standing problem in relativistic quantum mechanics
(see, for example, Refs. \cite{RefX,RefXX,RefY,RefYY}). It remains
to be seen whether any of these probability current densities, in
(1+1) dimensions, could also completely characterize a 1D KFGM particle
that lies within an interval. On the other hand, the problems of the
distinct types of energy densities and energy current densities that
can be introduced into nonrelativistic quantum mechanics and their
physical feasibility have also been studied (see Refs. \cite{RefZ,RefZZ}
and some references therein). 

Finally, our results can be extended to the problem of a 1D KFGM particle
in the real line with a point excluded, for example, at $x=0$ (thus,
$\Omega=\mathbb{R}-\{0\}=\mathbb{R}^{+}\,\cup\,\mathbb{R}^{-}$).
The general set of pseudo self-adjoint BCs for this problem can be
obtained from those corresponding to the particle in the interval
$\Omega=[a,b]$ by identifying the ends of this interval with the
two sides of the hole, i.e., $x=a\rightarrow0+$ and $x=b\rightarrow0-$.
Notably, each confining BC for this system defines a distinct impenetrable
barrier at $x=0$; in fact, these BCs isolate the $\mathbb{R}^{+}$
region from the $\mathbb{R}^{-}$ region. In contrast, each nonconfining
BC defines a specific transparent or permeable barrier at $x=0$.
These barriers allow the transmission of the particle between the
$\mathbb{R}^{+}$ and $\mathbb{R}^{-}$ regions.

\section*{Conflicts of interest}

\noindent The authors declare no conflicts of interest.

\section*{Data Availability Statement}

\noindent No datasets were generated or analysed during the current
study

\section*{Authors' Contributions}

\noindent Conceptualization, T.K. and S.DeV.; methodology, S.DeV.;
formal analysis, T.K. and S.DeV.; investigation, T.K. and S.DeV.;
resources, T.K. and S.DeV.; writing--original draft preparation, S.DeV.;
writing--review and editing, T.K. and S.DeV.; supervision, S.DeV.
Both authors have read and agreed to the current version of the manuscript.

\section*{\noindent Appendices}

\subsection*{\noindent A. On the local observables written in terms of two-component
wavefunctions}

\noindent The connections between the two-component wavefunctions
$\Psi$ and $\Phi$ and the respective one-component wavefunctions
$\psi$ and $\phi$ are given in Eq. (13). If we use $\Phi=\Psi$
in the relation given in Eq. (11) and in Eq. (10), with $\hat{\mathrm{h}}=\hat{\mathrm{h}}_{\mathrm{adj}}$,
it is clear that the term evaluated at the walls of the interval $\Omega$
in Eq. (10) vanishes, i.e., 
\[
\langle\langle\hat{\mathrm{h}}\Psi,\Psi\rangle\rangle=\langle\langle\Psi,\hat{\mathrm{h}}\Psi\rangle\rangle-\frac{\hbar}{\mathrm{i}}\left.\left[\, j\,\right]\right|_{a}^{b}=\langle\langle\Psi,\hat{\mathrm{h}}\Psi\rangle\rangle,\tag{A1}
\]
where 
\[
j=j(x,t)=\frac{\mathrm{i}\hbar}{2\mathrm{m}}\,\frac{1}{2}\left[\,\left((\hat{\tau}_{3}+\mathrm{i}\hat{\tau}_{2})\Psi_{x}\right)^{\dagger}(\hat{\tau}_{3}+\mathrm{i}\hat{\tau}_{2})\Psi-\left((\hat{\tau}_{3}+\mathrm{i}\hat{\tau}_{2})\Psi\right)^{\dagger}(\hat{\tau}_{3}+\mathrm{i}\hat{\tau}_{2})\Psi_{x}\,\right]\tag{A2}
\]
is precisely the usual current density given in Eq. (3). To go from
the expression in Eq. (A2) to that given in Eq. (3), the relation
between $\Psi$ and $\psi$ given in Eq. (13) must be used. In conclusion,
the result $j(b,t)=j(a,t)$ is a straightforward consequence of the
pseudo self-adjointness of $\hat{\mathrm{h}}$. On the other hand,
the formula given in Eq. (14) with $\Phi=\Psi$ and $\phi=\psi$ is
the following relation:
\[
\langle\langle\Psi,\Psi\rangle\rangle=\int_{\Omega}\mathrm{d}x\,\varrho(x,t),\tag{A3}
\]
where $\varrho$ is given in Eq. (2). If the definition of the scalar
product given in Eq. (9) is used in Eq. (A3), it is clear that 
\[
\varrho=\varrho(x,t)=\Psi^{\dagger}\hat{\tau}_{3}\Psi.\tag{A4}
\]

Similarly, if $\Phi=\Psi$ in the relation given in Eq. (20) with
$\hat{\mathrm{h}}=\hat{\mathrm{h}}_{\mathrm{adj}}$ (i.e., $\hat{\mathrm{h}}$
is a pseudo self-adjoint operator), then the term evaluated at the
boundaries of the interval given in Eq. (20) disappears, i.e., 
\[
\langle\langle\hat{\mathrm{h}}\Psi,\hat{\mathrm{E}}\Psi\rangle\rangle=\langle\langle\Psi,\hat{\mathrm{h}}\,\hat{\mathrm{E}}\Psi\rangle\rangle+\mathrm{i}\hbar\left.\left[\, j_{\mathrm{E}}\,\right]\right|_{a}^{b}=\langle\langle\Psi,\hat{\mathrm{h}}\,\hat{\mathrm{E}}\Psi\rangle\rangle\tag{A5}
\]
(see comments following Eqs. (20) and (26)), where 
\[
j_{\mathrm{E}}=j_{\mathrm{E}}(x,t)=-\frac{\hbar^{2}}{2\mathrm{m}}\,\frac{1}{2}\left[\,\left((\hat{\tau}_{3}+\mathrm{i}\hat{\tau}_{2})\Psi_{x}\right)^{\dagger}(\hat{\tau}_{3}+\mathrm{i}\hat{\tau}_{2})\dot{\Psi}-\left((\hat{\tau}_{3}+\mathrm{i}\hat{\tau}_{2})\Psi\right)^{\dagger}(\hat{\tau}_{3}+\mathrm{i}\hat{\tau}_{2})\dot{\Psi}_{x}\,\right]\tag{A6}
\]
is the energy current density in Eq. (26). If we use the relationship
between $\Psi$ and $\psi$ given in Eq. (13), we can obtain the expression
of $j_{\mathrm{E}}$ given in Eq. (26) from the expression in Eq.
(A6). The energy density $\varrho_{\mathrm{E}}$ can also be written
immediately in terms of $\Psi$. The formula given in Eq. (22) leads
us to the following relation:
\[
\langle\langle\Psi,\hat{\mathrm{E}}\Psi\rangle\rangle=\int_{\Omega}\mathrm{d}x\,\varrho_{\mathrm{E}},\tag{A7}
\]
where $\varrho_{\mathrm{E}}$ is given in Eq. (25). Using the scalar
product definition given in Eq. (9) in the latter formula, one can
write the energy density as follows:
\[
\varrho_{\mathrm{E}}=\varrho_{\mathrm{E}}(x,t)=\Psi^{\dagger}\hat{\tau}_{3}\,\hat{\mathrm{E}}\Psi.\tag{A8}
\]

As we know, $\varrho$ and $j$ vanish everywhere when a Majorana
condition is imposed on them. We can check this result again if we
use Eqs. (A4) and (A2), respectively. For example, for $\varrho$,
we can start by taking the complex conjugate of Eq. (A4), from which
we obtain the result $\varrho=\varrho^{*}$. We can then use a Majorana
condition, namely, $\Psi=\hat{\tau}_{1}\Psi^{*}$ or $\Psi=-\hat{\tau}_{1}\Psi^{*}$,
in Eq. (A4) \cite{RefE}, from which we obtain the relation $\varrho=-\varrho^{*}$.
Finally, the result $\varrho=0$ is obtained. This same procedure
can also be used to show that $j=0$. Likewise, $\varrho_{\mathrm{E}}$
and $j_{\mathrm{E}}$ in Eqs. (A8) and (A6), respectively, are real
quantities when they are calculated for a state satisfying a Majorana
condition. For example, for $\varrho_{\mathrm{E}}$, we can take the
complex conjugate of Eq. (A8) and use one of the Majorana conditions,
from which the result $\varrho_{\mathrm{E}}=\varrho_{\mathrm{E}}^{*}$
is obtained. The same procedure can be used to show that $j_{\mathrm{E}}=j_{\mathrm{E}}^{*}$. 

\subsection*{\noindent B. On the usual current density evaluated at the extremes
of an impenetrable interval}

\noindent The simple evaluation at $x=a$ of the usual current density
given in Eq. (3) allows us to write the following expression:
\[
j(x=a,t)=-\frac{\mathrm{i}\hbar}{2\mathrm{m}}\frac{1}{\lambda}\left[\,\psi^{*}(a,t)\,\lambda\psi_{x}(a,t)-\lambda\psi_{x}^{*}(a,t)\,\psi(a,t)\,\right].\tag{B1}
\]
The more general set of BCs for the solutions $\psi$ of the 1D KFG
equation given in Eqs. (17) and (18) can also take the form of the
family of BCs given in Eq. (45), i.e., 
\[
\left[\begin{array}{c}
\psi(b,t)\\
\lambda\psi_{x}(b,t)
\end{array}\right]=\frac{1}{\mathrm{m}_{1}+\mathrm{i}\,\mathrm{m}_{2}}\left[\begin{array}{cc}
\mathrm{m}_{3}+\sin(\mu) & -\mathrm{m}_{0}-\cos(\mu)\\
-\mathrm{m}_{0}+\cos(\mu) & -\mathrm{m}_{3}+\sin(\mu)
\end{array}\right]\left[\begin{array}{c}
\psi(a,t)\\
\lambda\psi_{x}(a,t)
\end{array}\right],\tag{B2}
\]
where $(\mathrm{m}_{0})^{2}+(\mathrm{m}_{1})^{2}+(\mathrm{m}_{2})^{2}+(\mathrm{m}_{3})^{2}=1$
and $\mu\in[0,\pi)$. Certainly, if one makes $\mathrm{m}_{2}=0$
in Eq. (B2), one obtains Eq. (45). From the matrix relation in Eq.
(B2), we can write $\lambda\psi_{x}(a,t)$ in terms of $\psi(a,t)$
and $\psi(b,t)$, namely, 
\[
\lambda\psi_{x}(a,t)=\left(\frac{\mathrm{m}_{3}+\sin(\mu)}{\mathrm{m}_{0}+\cos(\mu)}\right)\psi(a,t)-\left(\frac{\mathrm{m}_{1}+\mathrm{i}\,\mathrm{m}_{2}}{\mathrm{m}_{0}+\cos(\mu)}\right)\psi(b,t).\tag{B3}
\]
The complex conjugate of the latter relation is given by
\[
\lambda\psi_{x}^{*}(a,t)=\left(\frac{\mathrm{m}_{3}+\sin(\mu)}{\mathrm{m}_{0}+\cos(\mu)}\right)\psi^{*}(a,t)-\left(\frac{\mathrm{m}_{1}-\mathrm{i}\,\mathrm{m}_{2}}{\mathrm{m}_{0}+\cos(\mu)}\right)\psi^{*}(b,t).\tag{B4}
\]
Substituting Eqs. (B3) and (B4) into Eq. (B1), the formula that gives
the value of the usual current density at $x=a$ is immediately obtained,
namely, Eq. (19). 

The most general subset of confining BCs for a 1D KFG particle can
be obtained by making $\mathrm{m}_{1}=\mathrm{m}_{2}=0$ in the most
general family of BCs given in Eq. (17) \cite{RefK}. The result obtained
is precisely the two-parameter subset of BCs given in Eq. (65). From
the latter equation, we can write $\lambda\psi_{x}(a,t)$ in terms
of $\psi(a,t)$, namely, 
\[
\lambda\psi_{x}(a,t)=-\left[\frac{1-\mathrm{e}^{\mathrm{i}\,\mu}(\mathrm{m}_{0}+\mathrm{i}\,\mathrm{m}_{3})}{\mathrm{i}+\mathrm{i}\,\mathrm{e}^{\mathrm{i}\,\mu}(\mathrm{m}_{0}+\mathrm{i}\,\mathrm{m}_{3})}\right]\psi(a,t).\tag{B5}
\]
The complex conjugate of the latter relation is given by 
\[
\lambda\psi_{x}^{*}(a,t)=-\left[\frac{1-\mathrm{e}^{\mathrm{i}\,\mu}(\mathrm{m}_{0}+\mathrm{i}\,\mathrm{m}_{3})}{\mathrm{i}+\mathrm{i}\,\mathrm{e}^{\mathrm{i}\,\mu}(\mathrm{m}_{0}+\mathrm{i}\,\mathrm{m}_{3})}\right]\psi^{*}(a,t)\tag{B6}
\]
(we recall that the quantity inside the brackets in Eqs. (B5) and
(B6) is real). Substituting Eqs. (B5) and (B6) into Eq. (B1), we immediately
obtain the result
\[
j(x=a,t)=0,\tag{B7}
\]
as expected.

\subsection*{\noindent C. On the components of the standard energy-momentum tensor
in (1+1) dimensions}

\noindent The Lagrangian density for this system is given by
\[
\mathcal{L}=\frac{\hbar^{2}}{2\mathrm{m}}\left[\, g^{\mu\nu}(\partial_{\mu}\psi^{*})(\partial_{\nu}\psi)-\frac{\mathrm{m}^{2}c^{2}}{\hbar^{2}}\psi^{*}\psi-\frac{2\mathrm{m}S}{\hbar^{2}}\psi^{*}\psi\,\right],\tag{C1}
\]
where $\mu,\nu,\ldots=0,1.$, $g^{\mu\nu}=\mathrm{diag}(1,-1)$, $\partial_{0}=\partial/\partial(ct)$
and $\partial_{1}=\partial/\partial x$, as usual \cite{RefP}. The
Lagrangian density depends on $\psi$, $\partial_{\mu}\psi\equiv\partial\psi/\partial x^{\mu}$,
and $\psi^{*}$, $\partial_{\nu}\psi^{*}\equiv\partial\psi^{*}/\partial x^{\nu}$;
then, from the Euler-Lagrange equations
\[
\partial_{\alpha}\left[\frac{\partial\mathcal{L}}{\partial(\partial_{\alpha}\psi)}\right]-\frac{\partial\mathcal{L}}{\partial\psi}=0\quad\mathrm{and}\quad\partial_{\alpha}\left[\frac{\partial\mathcal{L}}{\partial(\partial_{\alpha}\psi^{*})}\right]-\frac{\partial\mathcal{L}}{\partial\psi^{*}}=0,\tag{C2}
\]
we obtain the complex conjugate of the 1D KFG equation in Eq. (1)
\[
\hbar^{2}c^{2}\left(\, g^{\mu\alpha}\partial_{\alpha}\partial_{\mu}+\frac{\mathrm{m}^{2}c^{2}}{\hbar^{2}}+\frac{2\mathrm{m}S}{\hbar^{2}}\,\right)\psi^{*}=0\tag{C3}
\]
(with $\mathrm{i}\hbar c\,\partial_{0}=\hat{\mathrm{E}}$ and $-\mathrm{i}\hbar\,\partial_{1}=\hat{\mathrm{p}}$)
and the 1D KFG equation in Eq. (1)
\[
\hbar^{2}c^{2}\left(\, g^{\mu\alpha}\partial_{\alpha}\partial_{\mu}+\frac{\mathrm{m}^{2}c^{2}}{\hbar^{2}}+\frac{2\mathrm{m}S}{\hbar^{2}}\,\right)\psi=0,\tag{C4}
\]
respectively. 

The commonly used energy-momentum tensor is given by \cite{RefP}
\[
T_{\mu}^{\;\;\nu}=(\partial_{\mu}\psi)\frac{\partial\mathcal{L}}{\partial(\partial_{\nu}\psi)}+(\partial_{\mu}\psi^{*})\frac{\partial\mathcal{L}}{\partial(\partial_{\nu}\psi^{*})}-\mathcal{L}\, g_{\mu}^{\;\;\nu}=\frac{\hbar^{2}}{2\mathrm{m}}\left[\,(\partial^{\nu}\psi^{*})\,(\partial_{\mu}\psi)+(\partial_{\mu}\psi^{*})\,(\partial^{\nu}\psi)\,\right]-\mathcal{L}\, g_{\mu}^{\;\;\nu},\tag{C5}
\]
whose components work out to be (a) $T_{0}^{\;\;0}=T_{\;\;0}^{0}$,
where $T_{\;\;0}^{0}$ is given in Eq. (31). (b) $T_{1}^{\;\;0}=-T^{10}=-T_{\;\;0}^{1}$,
where $T_{\;\;0}^{1}$ can be immediately obtained from Eq. (33).
(c) $T_{0}^{\;\;1}=T^{01}=T^{10}=+T_{\;\;0}^{1}$ (see Eq. (33)).
(d) $T_{1}^{\;\;1}=T_{\;\;1}^{1}$, where 
\[
T_{\;\;1}^{1}=\frac{1}{2\mathrm{m}c^{2}}\left[\,+(\hat{\mathrm{E}}\psi^{*})(\hat{\mathrm{E}}\psi)+(c\hat{\mathrm{p}}\psi^{*})(c\hat{\mathrm{p}}\psi)+(\mathrm{m}c^{2})^{2}\,\psi^{*}\psi+2\,\mathrm{m}c^{2}S\,\psi^{*}\psi\,\right].\tag{C6}
\]
The pair of components $\{T_{\;\;0}^{0},cT_{\;\;0}^{1}\}$ satisfies
the relation $\partial_{\mu}T_{\;\;0}^{\mu}=(\partial_{0}S)\psi^{*}\psi$;
similarly, the pair $\{T_{\;\;1}^{0},cT_{\;\;1}^{1}\}$ satisfies
$\partial_{\mu}T_{\;\;1}^{\mu}=(\partial_{1}S)\psi^{*}\psi$. Combining
the latter two formulas, we can write the following relation:
\[
\partial_{\mu}T_{\;\;\nu}^{\mu}=(\partial_{\nu}S)\psi^{*}\psi.\tag{C7}
\]
If the Lorentz scalar potential does not depend explicitly on time,
we can obtain a conservation law (i.e., $\int_{\Omega}\mathrm{d}x\, T_{\;\;0}^{0}=\mathrm{const}$)
only when the relation $\left.\left[\, c\, T_{\;\;0}^{1}\,\right]\right|_{a}^{b}=0$
is verified (as we have seen, the latter result depends on the BCs).
Similarly, if $S$ does not depend explicitly on the position, we
can obtain a conservation law (i.e., $\int_{\Omega}\mathrm{d}x\, T_{\;\;1}^{0}=\mathrm{const}$)
only when the relation $\left.\left[\, c\, T_{\;\;1}^{1}\,\right]\right|_{a}^{b}=0$
is verified. In this article, we do not discuss the latter conservation
law. 

\subsection*{D. On the energy densities $\varrho_{\mathrm{E}}$ and $j_{\mathrm{E}}$}

\noindent As is well known, the usual two-vector current density is
given by 
\[
j^{\mu}=j^{\mu}(x,t)=\frac{\mathrm{i}\hbar}{2\mathrm{m}}[\,\psi^{*}(\partial^{\mu}\psi)-(\partial^{\mu}\psi^{*})\,\psi\,],\tag{D1}
\]
where $j^{0}=c\varrho$ and $j^{1}=j$ (see Eqs. (2) and (3)). The
continuity equation takes the usual form $\partial_{\mu}j^{\mu}=0$
\cite{RefP}. We introduce the following second-rank Lorentz tensor:
\[
K_{\;\;\nu}^{\mu}=K_{\;\;\nu}^{\mu}(x,t)=\frac{\mathrm{i}\hbar}{2\mathrm{m}}\left[\,\psi^{*}\left(\partial^{\mu}(\mathrm{i}\hbar\,\partial_{\nu}\psi)\right)-(\partial^{\mu}\psi^{*})\,(\mathrm{i}\hbar\,\partial_{\nu}\psi)\,\right].\tag{D2}
\]
This tensor is an unusual energy-momentum tensor. Tensors $K_{\;\;\nu}^{\mu}$
and $T_{\;\;\nu}^{\mu}$ in Eq. (C5) are connected via the following
relation: 
\[
K_{\;\;\nu}^{\mu}=T_{\;\;\nu}^{\mu}+\mathcal{L}\, g_{\;\;\nu}^{\mu}-\frac{\hbar^{2}}{2\mathrm{m}}\,\partial_{\nu}\left(\psi^{*}(\partial^{\mu}\psi)\right).\tag{D3}
\]
Using Eqs. (C1), (C3) and (C4), the following relation can be demonstrated:
\[
\partial_{\mu}K_{\;\;\nu}^{\mu}=\partial_{\mu}T_{\;\;\nu}^{\mu}.\tag{D4}
\]
Now, using the result in Eq. (C7), we obtain the result
\[
\partial_{\mu}K_{\;\;\nu}^{\mu}=(\partial_{\nu}S)\psi^{*}\psi,\tag{D5}
\]
that is, $\partial_{\mu}K_{\;\;0}^{\mu}=0$ when $\dot{S}=0$. Precisely,
we have that $K_{\;\;0}^{0}=\varrho_{\mathrm{E}}$ and $K_{\;\;0}^{1}=j_{\mathrm{E}}/c$
(also $K^{01}=K^{10}$ when $\psi$ satisfies a Majorana condition,
i.e., in this case, $K^{\mu\nu}$ is a symmetric tensor). Then, we
obtain the conservation law $\int_{\Omega}\mathrm{d}x\, K_{\;\;0}^{0}=\int_{\Omega}\mathrm{d}x\,\varrho_{\mathrm{E}}=\langle\langle\Psi,\hat{\mathrm{E}}\Psi\rangle\rangle=\langle\langle\Psi,\hat{\mathrm{h}}\Psi\rangle\rangle=\mathrm{const}$
only when the relation $\left.\left[\, c\, K_{\;\;0}^{1}\,\right]\right|_{a}^{b}=\left.\left[\, j_{\mathrm{E}}\,\right]\right|_{a}^{b}=0$
is verified. As we have seen, the latter relation is a direct consequence
of the pseudo self-adjointness of the Hamiltonian $\hat{\mathrm{h}}$
(see Eq. (27)). Similarly, if $S$ does not depend explicitly on the
position, we have another conservation law $\partial_{\mu}K_{\;\;1}^{\mu}=0$.
In this article, we do not discuss this conservation law.

\end{document}